\documentclass[%
 reprint,
 amsmath,amssymb,
pra,
]{revtex4-2}

\usepackage{graphicx}
\usepackage{dcolumn}
\usepackage{bm}

\usepackage{physics}
\usepackage{mathtools}

\usepackage{hyperref}
\usepackage{color}
\definecolor{myblue}{RGB}{0,0,255}
\hypersetup{
    colorlinks,
    citecolor=myblue,
    linkcolor=myblue,
    urlcolor=myblue
}

\usepackage{color}
\usepackage{soul}
\definecolor{mycyan}{RGB}{73,8,138}
\definecolor{mypink}{cmyk}{0,0.7808,0.4429,0.1412}
\definecolor{mygreen}{RGB}{0,102,51}

\begin{document}

\title{Fundamental trade-off relation in probabilistic entanglement generation}
\author{Yuanbo Chen}
\email{chen@biom.t.u-tokyo.ac.jp}
\author{Yoshihiko Hasegawa}
\email{hasegawa@biom.t.u-tokyo.ac.jp}
\affiliation{
Graduate School of Information Science and Technology, 
The University of Tokyo, Tokyo 113-8656, Japan
}


\begin{abstract}
We investigate the generation of entanglement between two non-interacting systems by synthesizing a new quantum process from the superposition of distinct processes characterized by local-only operations. Our analysis leads to the derivation of a universal trade-off relation, $P_{\text{succ}}(1+\mathcal{C})\le1$, that fundamentally bounds the success probability ($P_{\text{succ}}$) and the generated entanglement (concurrence $\mathcal{C}$). The derivation of this  trade-off relation is inspired by indefinite causal order, but applies for a broader class of quantum processes. Next, we show that the mathematical structure of this bound predicts the existence of a ``quasi-deterministic'' mode of operation, a surprising phenomenon which we then confirm with concrete entanglement generation protocols, where a maximally entangled state is guaranteed to be produced. In this mode of operation, both outcomes of the post-selection measurement on the auxiliary control system result in a maximally entangled state of the target system.  Furthermore, we demonstrate how this general principle can be realized using a quantum switch, which leverages an indefinite causal order as a physical resource, and explore the rich variety of dynamical behaviors governed by the universal trade-off. Our results establish a general principle for entanglement generation with superposition of quantum processes and introduce a novel way of controlling entanglement generation.
\end{abstract}
\maketitle

\section{Introduction}
In the field of quantum information science, entanglement~\cite{RevModPhys.81.865, RevModPhys.86.419, PhysRevA.109.012404,PhysRevLett.115.250401} stands as a foundational element. This unique quantum form of correlation is widely recognized as an essential resource that enables quantum technologies to gain advantage over their classical counterparts in diverse tasks~\cite{RevModPhys.92.025002,Dutta_2023, Primaatmaja2023securityofdevice, Simon2017}. These applications range from quantum teleportation~\cite{Bouwmeester1997, PhysRevLett.83.5158, PhysRevLett.70.1895, Nielsen:2011:QCQI}, high-precision quantum metrology~\cite{PhysRevLett.96.010401,doi:10.1126/science.1104149} to fundamentally secure quantum cryptography~\cite{PhysRevLett.84.4729, PhysRevLett.67.661}, quantum network and internet~\cite{Kimble2008, doi:10.1126/science.aan3211, PhysRevLett.124.110501} to name a few. Given its importance, developing effective and controllable methods for creating this non-classical correlation is therefore a crucial and highly active area of research.

The conventional physical paradigms share a common feature that interactions between quantum systems are the primary mechanism for generating entanglement~\cite{RevModPhys.91.021001}. For example, in quantum many-body systems where strong couplings are present, a vast web of entanglement may be created out of the complicated dynamics. Exploiting interactions has consequently inspired numerous methods~\cite{PhysRevA.59.2468}, for generating entanglement in different contexts~\cite{RevModPhys.84.777}, relying on for example, direct interactions, the use of mediator systems~\cite{PhysRevLett.89.277901, PhysRevResearch.5.043295}, or in a remote way~\cite{RieraSabat2023remotelycontrolled}. Nevertheless, this reliance on interaction as a prerequisite raises a profound question: one is curious about whether entanglement can be generated when direct interaction between the two quantum systems is absent. 

In this work, our investigation of this kind of protocol leads to the derivation of a fundamental trade-off relation between the success probability of the protocol and the amount of entanglement it can generate. We demonstrate that this inequality is, in fact, a general principle that governs a broad class of superposition of quantum processes built on non-interacting Hamiltonians, including but not limited to those based on indefinite causal order (ICO)~\cite{Oreshkov:2012:NCOMMS, Goswami:2018:PRL, Rubino:2017:SCIADV, Procopio:2015:NCOMMS} or coherent control schemes. Furthermore, we show that an analysis of this universal bound predicts the existence of a ``quasi-deterministic'' mode of operation, a surprising phenomenon which we then confirm with concrete protocols, where a maximally entangled state is guaranteed to be produced. In parallel with our study, other recent works have proposed specific protocols for creating various classes of entangled states. These alternative approaches are typically formulated within a gate-based framework, restricting their analysis to the scope of ICO, and in some cases, utilizing a preshared maximally entangled state as a key resource~\cite{Koudia23, LiuPRAPPL25}.

The remainder of this paper is structured as follows. In the next section, we introduce the quantum switch as a physical resource capable of realizing an ICO. We then detail the construction of our entanglement generation protocol in Sec.~\ref{sec:3}, outlining the core assumption of non-interacting Hamiltonians and the conditions for the initial states. Sec.~\ref{sec:4} presents our central result, that is the derivation of a fundamental trade-off relation that connects the success probability of the protocol with the amount of generated entanglement. In Sec.~\ref{sec:5}, we demonstrate the rich variety of dynamical behaviors governed by this trade-off, including the discovery of a ``quasi-deterministic" mode where maximal entanglement is guaranteed. Finally, a self-contained proof of this universal bound is provided in Appendix~\ref{sec:proof}, with its extension to mixed states in Appendix~\ref{sec:mix_extension}. We also present a numerical study of the robustness of our bound in Appendix~\ref{sec:perturb}, while derivation details for specific numerical demonstrations are provided in Appendix~\ref{sec:der_2021}.
\begin{figure*}[htbp]
    \centering
    \includegraphics[width=0.99\textwidth]{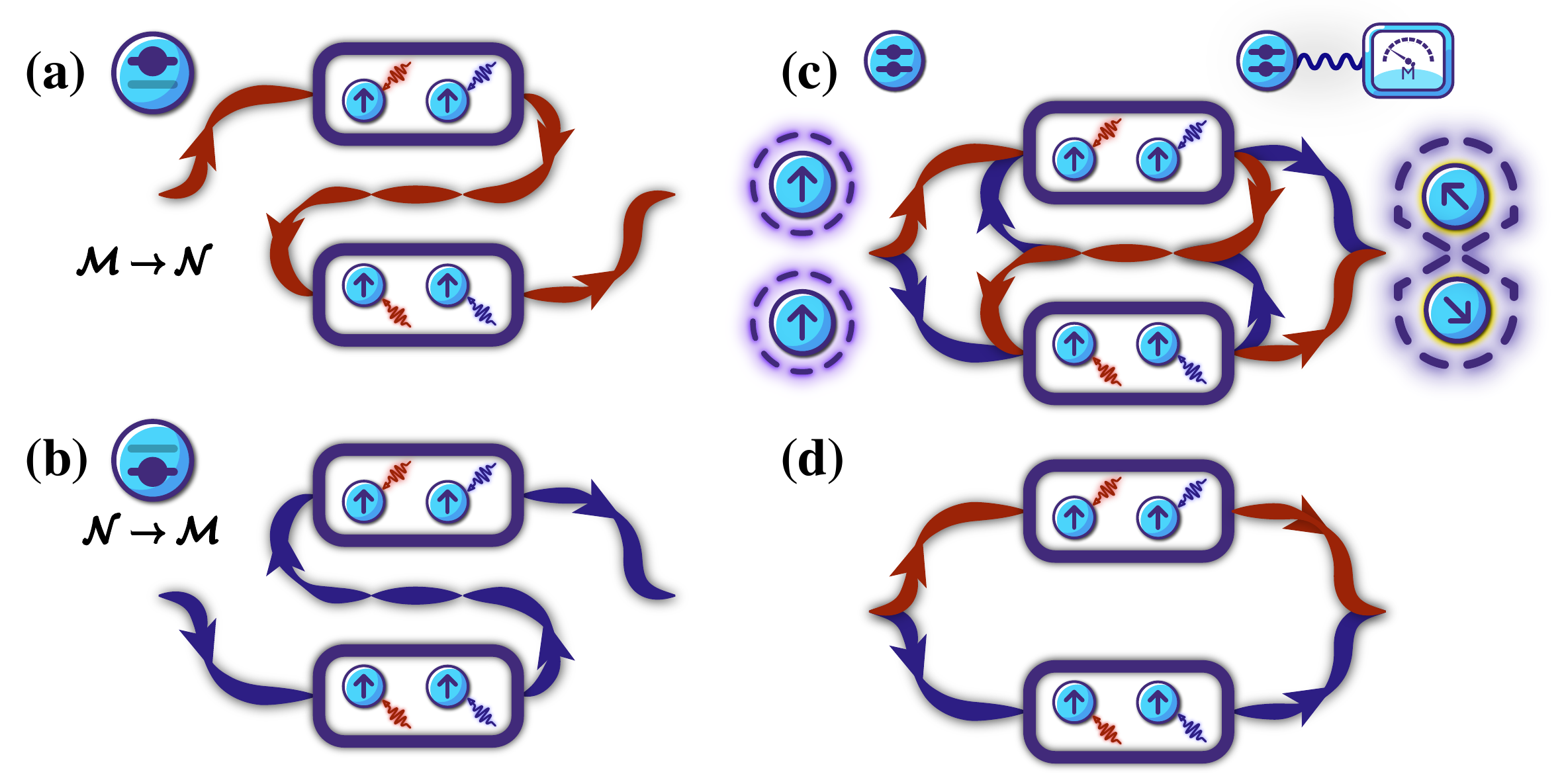}
    \caption{Schematic of the different architectures discussed in the work for entanglement generation. (a) When the control system (here a qubit) is in the state $\ket{0}^C$, the target qubits, $\mathcal{Q}^A$ and $\mathcal{Q}^B$, undergo local process $\mathcal{M}$ followed by process $\mathcal{N}$. (b) When the control system is in the orthogonal state $\ket{1}^C$, the reverse causal order, $\mathcal{N}$ followed by $\mathcal{M}$, is applied. (c) The quantum switch realizing ICO. If the target qubits are initialized in a separable state. The control system, prepared in a superposition, coherently controls the order of operations. A final measurement on the control system performs the post-selection. The resulting state of the target qubits can be entangled even though $\mathcal{Q}^A$ and $\mathcal{Q}^B$ do not directly interact. (d) A simplified diagram illustrating how the superposition of two distinct, non-interacting evolution paths gives rise to coherent control schemes.}
    \label{fig:epr}
\end{figure*}

\section{Quantum switch realizing novel quantum processes}
\label{sec:2}
 We begin by introducing the quantum switch~\cite{Chiribella:2013:PRA, Rubino:2021:PRR, Guerin:2016:PRL, Araujo:2014:PRL, Molitor2024, PhysRevResearch.5.023111}, and review necessary prerequisites to understand this novel physical resource. Recall that the evolution of a quantum system is dictated by the Schr\"{o}dinger equation, taking into account specific measurement processes, one could in principle find how the system evolves toward the future. Equivalently, an evolved state can be obtained by treating it as the output of the system passing a quantum channel, which is usually described by a completely positive and trace preserving (CPTP) map. 
 
 In addition, it is convenient to use the operator-sum representation for describing the action of a quantum process, in which case a set of operators called Kraus operators enables one to obtain the transformed state of a quantum system. Suppose that we have a channel $\mathcal{M}$ which is associated with a set of Kraus operators $\{K_i\}$. With the initial state $\rho$ (the time argument has been suppressed) of a system, the evolved state will then be $\widetilde{\rho}=\sum_iK_i\rho K_i^{\dagger}$, where the operators satisfy $\sum_iK_i^{\dagger}K_i=\mathbb{I}$, and an evolved state is marked with a tilde for the purpose of distinguishing from the initial state.

In cases where one encounters multiple processes, a fixed background causal structure is assumed even in the conventional setting of quantum mechanics. For instance, say we have two processes taking place in the space-time, one of the processes must exist in the causal past (or future) of the other. However, this assumption may not necessarily hold in quantum theory --- modifications should be made on the conventional notion of causality~\cite{Hardy_2007, Brukner2014,RRM+22}, if one takes into consideration the properties of both quantum mechanics and general relativity. In other words, nondeterminism of the quantum theory and the dynamical causal structure in general relativity may give rise to novel causal structures. What this implies is that one may expect the existence of processes whose causal order becomes \emph{indefinite} in a theory of quantum gravity, the main goal of which is to reconcile the two above-mentioned solid pillars of modern physics.

Processes where the causal order among events cannot be described by a definite sequence are referred to as possessing \emph{indefinite causal order}~\cite{PRXQuantum.5.010325, Zych2019}. Let us call two arbitrary processes $\mathcal{M}$ and $\mathcal{N}$ respectively. Especially in our consideration, $\mathcal{M}$ and $\mathcal{N}$ are modeled as two quantum channels. Depending on the way they locate in the space-time, $\mathcal{M}$ can exist in the causal past of $\mathcal{N}$, which we denote by $\mathcal{M}\circ\mathcal{N}$. Similarly, the composition of two channels in the reverse order $\mathcal{N}\circ\mathcal{M}$ means that $\mathcal{N}$ locates in the causal past of $\mathcal{M}$. In conventional scenarios, no other possibilities for their causal order are conceivable, however, ICO relaxes the constraint that any pair of events has to obey a fixed causal order. One can understand an ICO process by imagining that -- $\mathcal{M}\circ\mathcal{N}$, which corresponds to the process where $\mathcal{N}$ happens in the future of $\mathcal{M}$ is being in superposition with $\mathcal{N}\circ\mathcal{M}$. It is doubtful whether our physical world really allows for ICO happening rather than remaining just as an unrealistic model. In fact, regardless of appearing counter-intuitive, an ICO process can be realized through a novel physical resource known as quantum switch, which has been demonstrated to equipped with the ability of enhancing tasks in different contexts~\cite{chen2021, chen2022, PhysRevLett.131.240401, GHH+20, GCP+20, RRE+21,G12}.

A quantum switch is an example (refer to illustrations in Fig.~\ref{fig:epr}) which exhibits ICO, where an auxiliary system, i.e., a control system serves to control the order of events. In addition to the control system, another one is used as the target qubit to perform specific tasks. In order to illustrate how a quantum switch works, we examine its mathematical description in the Kraus decomposition representation,
\begin{align}
	\label{eq:def_qsw}\nonumber
	V_{ij}(t)=&\ketbra{1}{1}^C\otimes{K^m_i}(t/2){K^n_j}(t/2)\\
	+&\ketbra{0}{0}^C\otimes{K^n_j}(t/2){K^m_i}(t/2).
\end{align}
This expression mathematically defines the action of the quantum switch. The operator $V$ acts on the combined Hilbert space of the control and target systems, where the tensor product symbol $\otimes$ separates the operators acting on the control from those on the target. The terms $\{K_i^m\}$ and $\{K_i^n\}$ are the Kraus operator sets for the two channels $\mathcal{M}$ and $\mathcal{N}$, respectively. It is helpful to view the quantum switch as a higher-order resource that transforms input channels into a new process. The core mechanism can be interpreted by considering the initial state of the control system, $\rho_C=\ketbra{\psi}{\psi}^C$. When the control system is prepared in the definite state $\ketbra{0}{0}^C$, the resulting action on the target is the sequential application $\mathcal{N}\circ\mathcal{M}$. Conversely, preparing the control in the orthogonal state $\ketbra{1}{1}^C$ results in the reverse sequence, $\mathcal{M}\circ\mathcal{N}$.

The full system, consisting of the control system and the two target qubits, jointly evolves under the action of this quantum switch. Given an initial state prepared as $\rho_{CAB}\coloneqq\rho_C\otimes\rho_{AB}$, the system is mapped to the final state $\widetilde{\rho}_{CAB}=V\rho_{CAB}V^{\dagger}$. The specific initial states are crucial for our protocol: the control system $\rho_C$ is prepared in a balanced superposition, $\rho_C=\sum_{ij}(1/2)\ketbra{i}{j}^C$, which is necessary to activate the superposition of different causal paths. The target qubit pair is assumed to be prepared in a simple, separable state, $\rho_{AB}=\rho_A\otimes\rho_B$, which ensures that no entanglement is present in the initial state.

In addition to the state preparation and unitary evolution, the construction of our entanglement generating protocol requires a final measurement on the control system. This step, known as post-selection, is critical for our construction. For this protocol, we make the specific choice of using the basis $\ket{\pm}^C\coloneqq(\ket{0}^C\pm\ket{1}^C)/\sqrt2$ for this measurement. The final state of the target qubits $\mathcal{Q}^A$ and $\mathcal{Q}^B$ is therefore the key outcome of interest, as this state would manifest any entanglement generated by the protocol. Tracing out the degrees of freedom of the control system after the projective measurement yields the desired result, which can be formally expressed in the following way
\begin{align}
	\label{eq:def_rhopm}\nonumber
	\widetilde{\rho}_{AB}^{\pm}&=\text{tr}_C\big[\ketbra{\pm}^C\widetilde{\rho}_{CAB}\ketbra{\pm}^C\big]\\
	&=\sum_i\bra{i}^C\ketbra{\pm}^C\widetilde{\rho}_{CAB}\ketbra{\pm}^C\ket{i}^C,
\end{align}
where we denote two branches of outcome as $\widetilde{\rho}_{AB}^{\pm}$, which up to normalization is a density operator, coming with the probability of ${\rm{Tr}[\widetilde{\rho}_{AB}^{\pm}]}$. In the following, we first focus on the $\ket{-}^C$ outcome branch, while the link to its orthogonal post-selection outcome, i.e., that of $\ket{+}^C$ is discussed in the Appendix~\ref{sec:proof}.

\section{Construction of the protocol}
\label{sec:3}
We now specify the physical scenario, and the core assumptions about the dynamical aspect underlying our protocol. As we have stated earlier, the system under consideration consists of a pair of qubits, which are labeled as $\mathcal{Q}^A$ and $\mathcal{Q}^B$. A crucial assumption of our construction is that we are restricted to performing only \emph{local} operations on each of the two qubits. This is a critical constraint which equivalently states that no direct physical interaction is permitted to occur between $\mathcal{Q}^A$ and $\mathcal{Q}^B$. From the perspective of Hamiltonian dynamics, this no-interaction condition dictates that the joint Hamiltonian of the system, ${\mathcal{H}}^{AB}$, must be separable and is therefore simply the sum of the individual Hamiltonians:
\begin{equation}
\label{eq:def_sepH}
{\mathcal{H}}^{AB}={\mathcal{H}}^{A}\otimes\mathbb{I}^B+\mathbb{I}^A\otimes{\mathcal{H}}^{B},
\end{equation}
where ${\mathcal{H}}^{A(B)}$ denotes the local Hamiltonian operator for the corresponding qubit. We explicitly write out the tensor product structure with the identity operator $\mathbb{I}$ to emphasize the Hilbert space upon which each operator acts. It can be quickly observed that any unitary evolution generated by a Hamiltonian of this non-interacting form is incapable of creating quantum correlations. That is, if the pair of qubits is initially in a product, or separable, state, it will remain separable at all future times under such dynamics.

\subsection{Initial state considerations}
Having established the constraints on the generator of system dynamics, we now turn our attention to the choice of initial states for the qubits $\mathcal{Q}^A$ and $\mathcal{Q}^B$. One motivation for this work is to explore the possibility of generating entanglement, and particularly maximally entangled states, using ICO with a non-interacting Hamiltonian. Since the canonical example of a maximally entangled state for a two-qubit system is the pure EPR pair, it is both physically and mathematically natural to focus our analysis on initial configurations that can evolve into pure states. We therefore dedicate this section to discussing the conditions that the initial states must satisfy to achieve this goal.

Let us consider an arbitrary two-qubit initial state $\rho_{AB}$. When subjected to the quantum switch protocol outlined in the previous section, this state evolves into a post-selected state $\widetilde{\rho}_{AB}^{\pm}$. For our analysis, we will focus on the case where the evolution within each path of the quantum switch is governed by a closed Hamiltonian system. In this scenario, the quantum channels are described not by a sum of Kraus operators, but by a single unitary operator each. The Kraus operator set for a given channel thus consists of only one element, the unitary time evolution operator itself. This simplification allows us to write the explicit form of the post-selected state $\widetilde{\rho}_{AB}^{-}$ (corresponding to the $\ket{-}^C$ outcome) in terms of these unitary propagators:
\begin{align}
\label{eq:deltaUrho}
	\widetilde{\rho}_{AB}^{-} = &\frac{1}{4} (U_\mathcal{M}U_\mathcal{N} - U_\mathcal{N}U_\mathcal{M})\rho_{AB}(U_\mathcal{M}U_\mathcal{N} - U_\mathcal{N}U_\mathcal{M})^{\dagger},
\end{align}
where $U_\mathcal{M}$ and $U_\mathcal{N}$ are arbitrary local unitary operators acting on the Hilbert space of $\mathcal{Q}^A$ and $\mathcal{Q}^B$. The action is described by the commutator-like superoperator $(U_\mathcal{M}U_\mathcal{N} - U_\mathcal{N}U_\mathcal{M})$ acting on the initial state.

To examine the transformed state, we use the common metric of purity, which is defined as $ \mathcal{P}(\rho) \coloneqq \text{Tr}[\rho^2] $. The post-selected state $ \widetilde{\rho}_{AB}^{-} $ from Eq.~\eqref{eq:deltaUrho} can be written in the more compact form $\widetilde{\rho}_{AB}^{-} = \frac{1}{4} \Delta U_\mathcal{MN}\rho_{AB}\Delta U_\mathcal{MN}^{\dagger}$, with the operator $\Delta U_\mathcal{MN}$ defined as
\begin{align}
	\Delta U_\mathcal{MN} = U_\mathcal{M}U_\mathcal{N} - U_\mathcal{N}U_\mathcal{M}.
\end{align}

We now seek to establish the conditions under which our protocol can yield a pure quantum state. To begin, let us assume for the moment that the initial state of the two-qubit density operator, $ \rho_{AB} $, is arbitrary. This state evolves according to the ICO dynamics, which superposes the two unitary processes $ U_\mathcal{M}U_\mathcal{N} \circ U_\mathcal{N}U_\mathcal{M} $ and its reverse. To determine when the final state is pure, we analyze its purity, $\mathcal{P}(\widetilde{\rho}_{AB}^{-})$. For a normalized state, the purity is defined as $\mathcal{P}(\rho) = \text{Tr}(\rho^2)$, and it equals 1 if and only if the state is pure. In our case, the normalized post-selected state is $\widetilde{\rho}_{AB}^{-}/P_{\text{post}}$, so its purity is given by:
\begin{align}\nonumber
	\mathcal{P}(\widetilde{\rho}_{AB}^{-})&=\text{Tr}\left[\left(\frac{\widetilde{\rho}_{AB}^{-}}{P_{\text{post}}}\right)^2\right] \\\nonumber
	&= \frac{\text{Tr}[(\widetilde{\rho}_{AB}^{-})^2]}{P_{\text{post}}^2} \\
	&=\frac{\text{Tr}[(\Delta U_\mathcal{MN}\rho_{AB}\Delta U_\mathcal{MN}^{\dagger})^2]}{P_{\text{post}}^2}.
\end{align}  
To further examine this expression, we can assess the rank structure of the quantum state. By definition, any density operator such as $ \widetilde{\rho}_{AB}^{-} $ is positive semi-definite. Its trace, which corresponds to the success probability of the post-selection, can be expressed as the sum of its non-negative eigenvalues $\mu_k$:
\begin{align}
	P_{\text{post}}=\text{Tr}(\widetilde{\rho}_{AB}^{-}) = \sum_k \mu_k.
\end{align}  
Similarly, the trace of its square is the sum of the squared eigenvalues:
\begin{align}
	\text{Tr}[(\widetilde{\rho}_{AB}^{-})^2] = \sum_k \mu_k^2.
\end{align}
From these fundamental relations, it is clear that the purity satisfies that
\begin{align}
	\mathcal{P}(\widetilde{\rho}_{AB}^{-}) = \frac{(\sum_k \mu_k^2)}{(\sum_j \mu_j)^2}\leq 1
\end{align}

This inequality becomes equality, i.e., $\mathcal{P}(\widetilde{\rho}_{AB}^{-})$ is equal to 1 if and only if just one eigenvalues among $\mu_k$ is 1, which is the defining condition for a rank-1 operator. Therefore, for our protocol to produce a pure state, the post-selected state $\widetilde{\rho}_{AB}^{-}$ must be of rank-1. This in turn implies that the unnormalized operator $ \Delta U_\mathcal{MN}\rho_{AB}\Delta U_\mathcal{MN}^{\dagger} $ must also be of rank-1.

The preceding analysis establishes that the final state $\widetilde{\rho}_{AB}^{-}$ is pure if and only if the operator $\Delta U_\mathcal{MN}\rho_{AB}\Delta U_\mathcal{MN}^{\dagger}$ is rank-1. It is therefore critical to determine the conditions on the initial state $ \rho_{AB} $ that ensure this rank-1 property is satisfied. The answer, as we will demonstrate, depends on the invertibility of the operator $\Delta U_\mathcal{MN}$.

In the case where $ \Delta U_\mathcal{MN}$ is an invertible operator, the transformation preserves the rank of the initial density matrix. This is expressed by the relation:
\begin{align}
	\text{rank}(\Delta U_\mathcal{MN}\rho_{AB}\Delta U_\mathcal{MN}^{\dagger}) = \text{rank}(\rho_{AB}).
\end{align}  
The invertibility of $ \Delta U_\mathcal{MN} $ ensures that the support of $\rho_{AB}$ is mapped to a space of the same dimension. Consequently, for the final state $\widetilde{\rho}_{AB}^{-} $ to be rank-1, the initial state $ \rho_{AB} $ must itself be rank-1, which is the definition of a pure state. This means that if one starts with any mixed state (where $\text{rank}(\rho_{AB}) > 1$), the final state will also be mixed, and the goal of producing a pure state via the ICO dynamics will not be achieved in this scenario.

Conversely, we consider the special case where $ \Delta U_\mathcal{MN} $ is not invertible, which occurs if this commutator-like operator has a nontrivial kernel. In this situation, it is possible for a mixed initial state $ \rho_{AB} $ to be mapped to a rank-1 final state, $\widetilde{\rho}_{AB}^{-}$, provided that the support of $\rho_{AB}$ lies entirely within the kernel of $\Delta U_\mathcal{MN}$. However, such scenarios are not generic and would require fine-tuned symmetries or specific degeneracies in the underlying unitary dynamics of $U_\mathcal{M}U_\mathcal{N} $ and $U_\mathcal{N}U_\mathcal{M} $. In the absence of these specific conditions, which we explicitly exclude in our following analysis, the requirement for $\widetilde{\rho}_{AB}^{-}$ to be a rank-1 operator enforces that the initial state $ \rho_{AB} $ must be pure. By restricting our focus to pure initial states, we can therefore guarantee that our protocol is capable of generating other pure states.

\section{A fundamental trade-off relation}
\label{sec:4}
\subsection{Concurrence-Probability Trade-off}  
The reliance of our protocol on post-selecting a specific measurement outcome upon the control system means that it is inherently probabilistic in nature. A crucial aspect of characterizing such a method is to understand not only the properties of the desired final state but also the likelihood of obtaining it. A key quantity of interest is therefore the amount of entanglement present in the state generated by the protocol. To this end, we employ the concurrence to quantify the degree of entanglement for the two-qubit system. Another central goal of this work is to elucidate the relationship between the success probability, $ P_{\text{succ}} $, and the final state concurrence, $ \mathcal{C}(\widetilde{\rho}_{AB}^-) $. This relationship is fundamental to understanding the limitations and capabilities of using ICO as a resource for entanglement generation. We therefore seek to derive a trade-off relation that connects these two quantities. Our analysis begins with an examination of the post-selected two-qubit state, $\widetilde{\rho}_{AB}^{-}$. For notational simplicity in the subsequent discussion, we will drop the superscript $^{-}$ and refer to the transformed state simply as $\widetilde{\rho}_{AB}$.

To formalize this analysis, we begin by specifying the initial state. Based on our conclusion from the previous section that pure initial states are required to generate pure final states, we consider a two-qubit initial state given by the pure, separable vector $\ket{\Psi}^{AB} = \ket{\psi}^A\otimes\ket{\phi}^B$. Here, $\ket{\psi}^A$ and $\ket{\phi}^B$ are arbitrary pure states for their respective qubits. The choice of a separable state ensures that any entanglement in the final state is generated solely by the protocol. The transformed state of this system, following the ICO dynamics and the tracing out of the control system, is:
\begin{align}
\widetilde{\rho}_A =\text{tr}_B\ketbra{\widetilde{\Psi}}^{AB}.
\end{align}  
For concurrence calculations, recall that the definition for a two-qubit state $ \rho_{AB} $, the standard formula reads:  
\begin{align}
\mathcal{C}(\rho_{AB}) = \max\left(0, \sqrt{\lambda_1} - \sqrt{\lambda_2} - \sqrt{\lambda_3} - \sqrt{\lambda_4}\right),
\end{align}  
where $ \lambda_i $ are the eigenvalues of $ \rho_{AB} (\sigma_y \otimes \sigma_y) \rho_{AB}^* (\sigma_y \otimes \sigma_y) $. For a pure (and normalized) state $ \rho_{AB}=\ketbra{\psi_{AB}} $, this simplifies to:  
\begin{align}
\mathcal{C}(\rho_{AB}) = \sqrt{2[1-\text{Tr}(\text{tr}_B\rho_{AB})^2]}.
\end{align}  
Specifically, under the assumption of the initial state being pure, and taking into consideration the normalization of a transformed state, the concurrence reads
\begin{align}
\label{eq:C_def_rhoa}
	\mathcal{C}(\widetilde{\Psi}_{AB}) = \sqrt{2[1-\text{Tr}(\widetilde{\rho}_{A}^2)/\text{Tr}(\widetilde{\rho}_{A})^2]},
\end{align}
where $\widetilde{\rho}_{A}^2=\big(\text{tr}_B\ketbra{\widetilde{\Psi}}^{AB}\big)^2$, and the quantity $\text{Tr}(\widetilde{\rho}_{A})^2$ is recognized as the square of success probability seeing this outcome after measurement on the control system.

The expression in Eq.~\eqref{eq:C_def_rhoa} provides a direct and practical way to quantify the degree of entanglement in the final two-qubit state, connecting it to the purity of the reduced state of one of the qubits. To move forward and demonstrate the main result of this paper, that is a quantitative relationship between the concurrence and the success probability, we write down the definition of the success probability.
\begin{figure*}[htbp]
    \centering
    \includegraphics[width=0.95\textwidth]{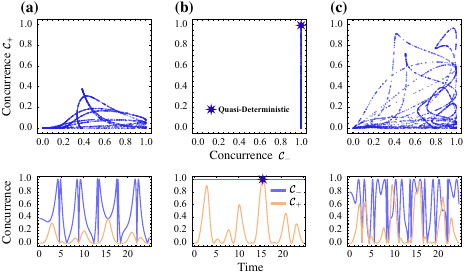}
    \caption{Entanglement distribution between the two post-selection branches for three different Hamiltonian dynamics. The top row shows the $(\mathcal{C}_-, \mathcal{C}_+)$ relationship generated by sampling the evolution at random times up to time $t_\text{short}=25$. The bottom row shows the corresponding short-time evolution of $\mathcal{C}_-$ (blue) and $\mathcal{C}_+$ (yellow). The dynamics clearly exhibit distinct patterns, especially the ``quasi-deterministic'' feature in the middle. This intriguing feature in (b) where two branches of evolution reach maximal concurrence simultaneously at a specific time, is highlighted by the star mark. The parameters for (a), (b) and (c) correspond to the Case 1, 2, and 3 defined in the text.}
    \label{fig:c_minus_vs_c_plus}
\end{figure*}
Recall that we have made the assumption about the initial state being pure, therefore, if a transformed state (without normalization) is 
\begin{align}
{{\ket{\widetilde{\Psi}}^{AB}}=\sum_{i,j}p_{ij}\ket{ij}^{AB}},
\end{align}  
where $p_{ij}\in\mathbb{C} $, and $\ket{ij}^{AB}=\ket{i}^A\otimes\ket{j}^B$ is taken as the basis for calculation. The success probability then has the simple form as below,
\begin{align}
P_{\text{succ}} = \sum_{ij}\abs{p_{ij}}^2.
\end{align}

A detailed proof for the following result that characterizes the relation between these two quantities, conditioned on the success probability can be found in Appendix~\ref{sec:proof}.
\begin{align}
\label{eq:???}
	\begin{cases}
		\mathcal{C}(\widetilde{\Psi}_{AB})\leq1, &\text{if }0 \leq P_{\text{succ}} \leq \frac{1}{2}\\
	\mathcal{C}(\widetilde{\Psi}_{AB})\leq {(1-P_{\text{succ}})}/{P_{\text{succ}}}, &\text{if }\frac{1}{2} \le P_{\text{succ}} \leq 1.
	\end{cases}
\end{align}
which essentially implies that $P_{\text{succ}}\leq 1/[1+\mathcal{C}(\widetilde{\Psi}_{AB})]$, and can be further cast into the form
\begin{align}
\label{eq:pc_trade_off}
P_{\text{succ}}[1+\mathcal{C}(\widetilde{\Psi}_{AB})]\leq1.
\end{align}
The inequality Eq.~\eqref{eq:pc_trade_off} the central result of our work, as it manifests a fundamental trade-off relation between the probability of success and the maximal achievable entanglement. This expression quantitatively captures the intuitive notion that generating states with a high degree of quantum correlation comes at a cost, as the desired concurrence $\mathcal{C}(\widetilde{\Psi}_{AB})$ approaches its maximum value, the upper bound on the success probability $P_{\text{succ}}$ becomes more restrictive.

\subsection{Comments on the Trade-off Relation}
It is crucial to note that while this trade-off relation $P_{\text{succ}}[1+\mathcal{C}(\widetilde{\Psi}_{AB})]\leq1$ is inspired by the specific structure of the ICO protocol, its mathematical proof (see Appendix~\ref{sec:proof}) does not restrict to the class of quantum processes that fits into the form that a quantum switch takes. The derivation relies only on the general properties of coherently interfering two valid quantum evolution paths and the non-interacting condition. Therefore, Eq.~\eqref{eq:pc_trade_off} represents a general bound applicable to a broader class of quantum processes, including conventional coherent control schemes. This structure is not exclusive to ICO, as illustrated in Fig.~\ref{fig:epr}(d), a conventional coherent control scheme where a control system conditionally applies one of two distinct unitaries ($U_0$ or $U_1$) to a target system also yields a mathematically identical form upon post-selection. By simply identifying these conditional unitaries with the two ordered operations, $U_0 \rightarrow U_\mathcal{M} U_\mathcal{N}$ and $U_1 \rightarrow U_\mathcal{N} U_\mathcal{M}$, the resulting operator on the target system becomes proportional to $(U_\mathcal{M} U_\mathcal{N} - U_\mathcal{N} U_\mathcal{M})$, thus recovering our central formalism.

Furthermore, the duality inherent in the proof, which reveals a symmetry between the ``difference'' (e.g., in the ICO case, that is $U_\mathcal{M}U_\mathcal{N} - U_\mathcal{N}U_\mathcal{M}$) and ``sum'' ($U_\mathcal{M}U_\mathcal{N} + U_\mathcal{N}U_\mathcal{M}$) resulting from the quantum superpositions of processes, implies that the same trade-off bound holds for the orthogonal $\ket{+}^C$ post-selection outcome. This raises the intriguing possibility of a ``quasi-deterministic'' mode of operation, where both branches might simultaneously produce a maximally entangled state, that is $\mathcal{C}_-=\mathcal{C}_+=1$, where the subscripts $\pm$ are used to explicitly distinguish a $\ket{-}^C$ outcome from the orthogonal $\ket{+}^C$ branch. In the following section, we explore this hypothesis and the consequences of the general bound by demonstrating entanglement generation in Hamiltonian dynamics.
\begin{figure*}[htbp]
    \centering
    \includegraphics[width=0.95\textwidth]{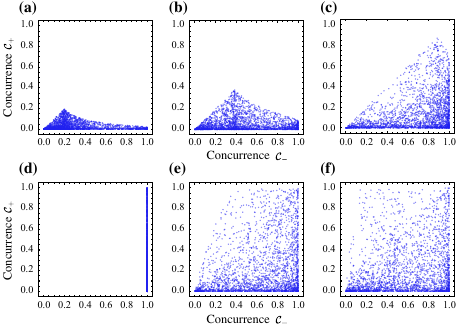}
    \caption{Long-time state space coverage in the $(\mathcal{C}_-, \mathcal{C}_+)$ plane for six Hamiltonian systems. Each one shows the distribution of points generated by uniformly sampling the evolution at 2500 of random times up to $T_\text{max}=1500$, illustrating the distinct long-time behavior of each system. The parameters for (a) to (f) are selected as $\omega_A=1.0, \omega_B=1.0, c_{N,B}^X=1.0$ being fixed, while $c_{M,A}^X$ takes the values of $0.1, 0.2, 0.6, 1.0, 2.0, 4.0$, respectively.}
    \label{fig:c_long_time}
\end{figure*}
Another particularly significant consequence of this trade-off becomes apparent when we consider the generation of a perfect EPR state, which is a maximally entangled state. For such a state, the concurrence is, by definition, maximal, with $\mathcal{C}=1$. Substituting this value into our trade-off inequality immediately reveals that the maximal probability to achieve this outcome is $1/2$.

One should be noted that that this trade-off relation is derived for an ideal coherent implementation of the protocol. In particular, it assumes that the coherent superposition between the two orders $U_\mathcal{M}U_\mathcal{N}$ and $U_\mathcal{N}U_\mathcal{M}$ is preserved. Therefore, decoherence mechanism that degrades this coherence~\cite{PhysRevLett.106.090502} will in general reduce the entanglement that can be generated. In the extreme case where no coherence remains, a quantum switch effectively becomes a ``classical switch'', enforcing the two orders contribute only \emph{incoherently}. This will make the protocol unable to generate entanglement, leading to $\mathcal{C}=0$, and the bound becomes trivial.

Furthermoe, in our protocol, $P_{\mathrm{succ}}$ is the \emph{intrinsic} Born probability of the desired control outcome in an ideal, fully coherent, lossless setting, and the trade-off constrains this quantity. In realistic photonic experiments, for instance, the observed heralding probability would be this intrinsic $P_{\mathrm{succ}}$ further reduced by factors like losses and detector inefficiencies.

\subsection{Extension to Mixed States and Robustness Analysis}
In preceding discussion, for simplicity, we have been focusing on pure initial states, which is natural in view of the goal of preparing pure entangled states. However, our trade-off is in fact more general. Using the pure-state result together with the convexity of concurrence, we prove in Appendix~\ref{sec:mix_extension} that the bound
\begin{equation}
P_{\mathrm{succ}}(\rho)\,\bigl[1+C\bigl(\rho_{\mathrm{out}}(\rho)\bigr)\bigr]\;\le\;1
\end{equation}
holds for \emph{any} separable two-qubit input state $\rho$. This shows that our trade-off is robust as long as the initial states are separable. This limitation on generating entanglement from separable mixed states parallels the thermodynamic unattainability principle, which forbids cooling a system to its ground state with finite resources. Here, the trade-off relation quantitatively captures the probabilistic cost of distilling quantum correlations from noisy initial conditions.

However, due to the fact that our trade-off relation is derived under two structural assumptions: (i) in each branch the target dynamics are strictly non-interacting, and (ii) the initial state of the two targets is separable, i.e., vanishing initial entanglement. Once either assumption is relaxed, however, weak direct interactions or weak pre-existing entanglement provide additional entangling resources, so the universal inequality $P_{\mathrm{succ}}(1+\mathcal{C}) \le 1$ need not hold exactly.

It is also of fundamental interest to understand how the trade-off behaves under small deviations from these ideal assumptions. In Appendix~\ref{sec:perturb} we present a numerical study in which we (a) introduce weak initial entanglement while keeping the dynamics non-interacting, and (b) add a small interaction term $\mathcal{H}_{\mathrm{int}}$ while starting from a separable input state. We observe that violations of the ideal trade-off are themselves perturbative in the size of these deviations. Developing a full analytic robustness theory, that is bounding the maximal violation in terms of $\lVert \mathcal{H}_{\mathrm{int}}\rVert$ and (or) the initial concurrence is an interesting direction for future work.
\section{Dynamical Consequences of the Trade-off Relation}
\label{sec:5}
The trade-off relation derived in Sec.~\ref{sec:4} defines the universal boundary for all possible entanglement dynamics within this class of superposition of quantum processes. However, the specific trajectory a system follows within this allowed region, and what patterns of entanglement emerge, is a highly non-trivial function of the underlying Hamiltonian dynamics.
Indeed, while our inequality defines the absolute boundary of what is possible, finding the specific Hamiltonian parameters that allow a dynamics to actually reach and saturate this bound is itself not trivial.

This non-trivial challenge of engineering entanglement-generating dynamics stands in contrast to the trivial case where no entanglement is produced at all. Such a trivial outcome occurs, for instance, if the two superposed evolution paths are identical, as their destructive interference yields a null result when focusing on the $\ket{-}^C$ outcome branch.

In this section, we demonstrate the rich variety of these physical behaviors. We first analyze the distribution of entanglement between the two post-selection branches, which leads to the discovery of the quasi-deterministic mode. We then provide a demonstration of the universal trade-off boundary that governs all such dynamics.
\begin{figure}[b]
    \includegraphics[width=0.49\textwidth]{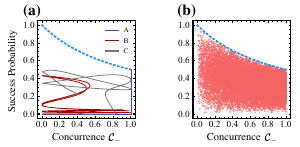}
    \caption{(a) Short-time dynamical trajectories in the $(\mathcal{C}_-, P_{\text{succ}})$ plane for the three exemplary Hamiltonian dynamics (A, B, C) defined as: $A:\omega_A=1.0, \omega_B=1.0, c_{M,A}^X=1.0, c_{N,B}^X=1.0$, up to time $t_\text{short}=20$, $B:\omega_A=1.0, \omega_B=5.0, c_{M,A}^X=1.0, c_{N,B}^X=1.0$, up to time $t_\text{short}=5$, and $C:\omega_A=1.0, \omega_B=1.0, c_{M,A}^X=1.0, c_{N,B}^X=1.5, c_{N,B}^Y=1.0$, up to time $t_\text{short}=5$. These confirms they evolve within the allowed region. (b) Demonstration of the trade-off boundary. Each point represents the outcome of a single trial with Haar-random local unitary operators. The distribution densely populates the entire allowed region and no point violates the theoretical boundary, that is the dashed blue line.}
    \label{fig:universality}
\end{figure}

To conduct our explorations, we consider a general physical model where the dynamics are governed by two process Hamiltonians, $\mathcal{H}_M$ and $\mathcal{H}_N$. The unitary time propagator is obtained as $U_\mathcal{M}(t)={\exp}(-i  \mathcal{H}_M t)$ and similarly for process $\mathcal{N}$ we have $U_\mathcal{N}(t)={\exp}(-i  \mathcal{H}_N t)$, where we set $\hbar=1$. Each of the two processes Hamiltonians is composed of a general, two-qubit internal Hamiltonian and four local auxiliary Hamiltonians:
\begin{align}
\label{eq:defHk}
    \mathcal{H}_k = (\omega_A \sigma_z^A \otimes \mathbb{I}^B + \mathbb{I}^A \otimes \omega_B \sigma_z^B) + \sum_{j \in \{A,B\}} \mathcal{H}_{\text{aux},k}^{(j)},
\end{align}
where $k\in\{M,N\}$, and each local auxiliary Hamiltonian is a linear combination of Pauli operators, e.g., $\mathcal{H}_{\text{aux},M}^{(A)} = c_{M,A}^X \sigma_x^A + c_{M,A}^Y \sigma_y^A + c_{M,A}^Z \sigma_z^A$. For example, the cases going to be presented in Fig~ \ref{fig:c_minus_vs_c_plus}
, we use the following Hamiltonian parameters:
\begin{itemize}
    \item \textbf{Case 1:} $\omega_A=1.0, \omega_B=1.0, c_{M,A}^X=0.2, c_{N,B}^X=1.0$.
    \item \textbf{Case 2 (Quasi-Deterministic):} $\omega_A=1.0, \omega_B=1.0, c_{M,A}^X=1.0, c_{N,B}^X=1.0$.
    \item \textbf{Case 3:} $\omega_A=1.0, \omega_B=1.0, c_{M,A}^X=2.0, c_{N,B}^X=1.0$.
\end{itemize}
At this point it is useful to clarify the relation between the general non-interacting Hamiltonian $ \mathcal{H}_{AB}$ in Eq.~\eqref{eq:def_sepH} and the process Hamiltonians $ \mathcal{H}_k$ introduced in Eq.~\eqref{eq:defHk}. First, Eq.~\eqref{eq:def_sepH} encodes the basic physical constraint of our protocol, i.e., all dynamics of the two-qubit target system must be generated by a separable, non-interacting Hamiltonian.
We then choose a concrete family of such Hamiltonians for the two processes. Thus Eq.~\eqref{eq:defHk} is a specific, parametrized realization of the general non-interacting form in Eq.~\eqref{eq:def_sepH}, used for illustrating the possible dynamics within the universal trade-off derived earlier.

In the numerical demonstrate presented below, we simulate the unitary evolution generated by the process Hamiltonians $\mathcal{H}_M$ and $\mathcal{H}_N$ defined in Eq.\eqref{eq:defHk}. For all results in Figs.~\ref{fig:c_minus_vs_c_plus}--\ref{fig:psucc_long_time}, the system is initialized in the separable product state $\ket{00}^{AB}$.

\subsection{Entanglement Distribution Between Post-Selection Branches}

We first investigate the relationship between the entanglement generated in the two post-selection branches, characterized by concurrences $\mathcal{C}_-$ and $\mathcal{C}_+$. In Fig.~\ref{fig:c_minus_vs_c_plus}, we show both the short-time evolution and the long-time evolution picture of entanglement in the $(\mathcal{C}_-, \mathcal{C}_+)$ plane for the three Hamiltonians defined above, i.e., parameters for Case 1, 2, and 3. The initial state condition for this demonstration and that follows is fixed to be $\ket{00}^{AB}$, as stated earlier.

We distinguish between two modes of analysis. Short-time dynamics (e.g. in Fig.~\ref{fig:c_minus_vs_c_plus}) correspond to continuous time evolution up to a fixed time $t_\text{short}$, revealing the immediate correlations between branches. Long-time behavior (e.g. in Fig.~\ref{fig:c_long_time}) is visualized via scatter plots obtained by sampling the dynamics at random times $t$ drawn uniformly from a large interval $[0,T_\text{max}]$ with, e.g., sample number = 2500 in Fig.~\ref{fig:c_long_time}, illustrating the asymptotic ``state-space'' coverage of a speicific Hamiltonian dynamics.

In the short-time evolution shown in Fig.~\ref{fig:c_minus_vs_c_plus}, the dynamics exhibit distinct patterns. Notably, for Case 2 in Fig.~\ref{fig:c_minus_vs_c_plus}(b), we observe a unique feature marked by the star symbol. At this specific instant, both post-selection branches simultaneously reach maximal concurrence. This implies that specific Hamiltonian symmetries can force the system into a ``quasi-deterministic'' mode where high entanglement is guaranteed regardless of the measurement outcome.

To understand the long-time behavior, we extend the time sampling for each case. The resulting long-time coverage plots in Fig.~\ref{fig:c_long_time} showcase the the ``ergodic" footprint of each Hamiltonian in the $(\mathcal{C}_-, \mathcal{C}_+)$ plane. For the specific simulation results shown in Fig.~\ref{fig:c_long_time}(a)--(f), we fix the parameters $\omega_A=\omega_B=1.0$ and $c_{N,B}^X=1.0$, while varying the coupling coefficient $c_{M,A}^X$ across the values $0.1, 0.2, 0.6, 1.0, 2.0,$ and $4.0$, respectively. The distribution of points generated by uniformly sampling the evolution at 2500 of random times up to $T_\text{max}=1500$.

\subsection{Universal Boundary Demonstration}
Having established the different patterns of entanglement distribution, we now provide a demonstration of the universal trade-off relation between success probability and concurrence. Fig.~\ref{fig:universality}(a) confirms that the short-time trajectories for our three exemplary Hamiltonians are also strictly governed by this same universal boundary. The Hamiltonian parameters for the labelled curves A, B, and C in Fig.~\ref{fig:universality}(a) are defined as: $A:\omega_A=1.0, \omega_B=1.0, c_{M,A}^X=1.0, c_{N,B}^X=1.0$, $B:\omega_A=1.0, \omega_B=5.0, c_{M,A}^X=1.0, c_{N,B}^X=1.0$, and $C:\omega_A=1.0, \omega_B=1.0, c_{M,A}^X=1.0, c_{N,B}^X=1.5, c_{N,B}^Y=1.0$. In Fig.~\ref{fig:universality}(b), we perform a demonstration that samples over the entire space of all possible local operations using Haar-random unitaries and initial separable states. The results densely populate the entire allowed region of the $(\mathcal{C}_-, P_{\text{succ}})$ plane, but no point violates the theoretical boundary derived in Sec.~\ref{sec:4}. This provides a concrete demonstration that our trade-off relation is a universal feature.

\begin{figure}[t]
    \centering
    \includegraphics[width=0.5\textwidth]{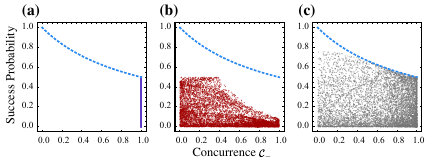}
    \caption{Long-time state space coverage in the $(\mathcal{C}_-, P_{\text{succ}})$ plane for the three examples in Fig.~\ref{fig:universality} (a). Each panel shows the distribution of points generated by sampling the evolution at random times up to $T_\text{max}=9000$, with 7000 uniform randomly sampled points. The distinct shapes of the distributions illustrate the long-time behavior of each system within the universal boundary of the dashed blue line. The Hamiltonian parameters for panel (a), (b), and (c) are defined as: $A:\omega_A=1.0, \omega_B=1.0, c_{M,A}^X=1.0, c_{N,B}^X=1.0$, $B:\omega_A=1.0, \omega_B=5.0, c_{M,A}^X=1.0, c_{N,B}^X=1.0$, and $C:\omega_A=1.0, \omega_B=1.0, c_{M,A}^X=1.0, c_{N,B}^X=1.5, c_{N,B}^Y=1.0$.}
    \label{fig:psucc_long_time}
\end{figure}

The long-time coverage in the $(\mathcal{C}_-, P_{\text{succ}})$ plane, shown in Fig.~\ref{fig:psucc_long_time}, further reinforces this conclusion. It illustrates the distinct long-time behavior of each system, showing how each explores its own unique sub-region of the universally allowed state space. The specific parameters for the simulation time and sample sizes are detailed in the captions of Fig.~\ref{fig:universality} and Fig.~\ref{fig:psucc_long_time}.

Fig.~\ref{fig:psucc_long_time} also displays the probabilistic cost of entanglement generation, highlighting that the ability to saturate the trade-off bound is non-trivial and governed by the choice of Hamiltonian parameters. While the inequality $P_{\mathrm{succ}}(1+\mathcal{C}) \le 1$ defines the universally allowed region, a specific system dynamics does not automatically explore this entire space. As shown by the comparison between the panels, different parameter choices confine the system to distinct subregions: generic dynamics, e.g., in Fig.~\ref{fig:psucc_long_time}(b) may be restricted to internal areas and can never reach the theoretical boundary, whereas the specific cases in Fig.~\ref{fig:psucc_long_time}(a) and (c) allow the density distribution to press explicitly against the physical limit. This demonstrates that the trade-off can be saturated only under the right dynamical conditions.

\subsection{A Concrete Example}
\label{sec:conc_exmp}
To provide a more concrete realization of our method, we now construct a protocol based on Hamiltonian dynamics to explicitly show how the final entangled state is determined by specific conditions on the Hamiltonian parameters and the choice of initial state. This demonstration is particularly useful as it reveals a simple, measurement-timing independent fashion that guarantees the generation of a maximally entangled state.

As has been defined in Eq.~\eqref{eq:defHk}, the term $(\omega_A \sigma_z^A \otimes \mathbb{I}^B + \mathbb{I}^A \otimes \omega_B \sigma_z^B)$ is the internal Hamiltonian for each of the qubits, which describes their free evolution in the absence of any auxiliary operations, that is ${\mathcal{H}}^{A(B)}_\text{Internal}=\omega_z^{A(B)}\sigma_z$, where the constant $\omega_z^{A(B)}$ is a parameter representing the energy splitting. Based on the insights we have gain from the observation in the preceding demonstrations of dynamical properties of various Hamiltonian dynamics, we focus specifically on the $\ket{-}^C$ branch. Furthermore, we set $\omega_z^{A(B)}=\omega_z$ to be identical while focusing on the situations where we have the auxiliary Hamiltonians have the form of only $ c_{M,A}^X$ and $ c_{N,B}^X$ taking non-trivial values.

\begin{table}[b]
\caption{\label{tab:table1}
Conditions for an ICO process to generate maximally entangled states. For a qubit pair, there are more than one way to get maximally entangled, thus we label the four possibilities with $\mathbb{A}$, $\mathbb{B}$, $\mathbb{C}$ and $\mathbb{D}$ respectively. In this table, we use a state vector instead of a density operator to represent a quantum state. All the states in the last column are reduced states corresponding to the $\ket{-}^C$ measurement results associated with the control system.}
\begin{ruledtabular}
\begin{tabular}{ccccc}
 &$\ket{\psi}^{AB}$&$\text{Hamiltonian Conditions}$ &$\widetilde{\ket{{\psi}}}^{AB,-}$&\\
\hline
$\mathbb{A}$& $\ket{00}$ & $c_{M,A}^X/c_{N,B}^X=1$ & $(\ket{01}-\ket{10})/\sqrt{2}$\\
$\mathbb{B}$& $\ket{00}$ & $c_{M,A}^X/c_{N,B}^X=-1$ & $(\ket{01}+\ket{10})/\sqrt{2}$\\
$\mathbb{C}$& $\ket{01}$ & $c_{M,A}^X/c_{M,B}^X=-1$ & $(\ket{00}-e^{i\phi}\ket{11})/\sqrt{2}$\footnotemark[1]\\
$\mathbb{D}$& $\ket{01}$ & $c_{M,A}^X/c_{M,B}^X=1$ & $(\ket{00}+e^{i\phi}\ket{11})/\sqrt{2}$\footnotemark[1]
\end{tabular}
\end{ruledtabular}
\footnotetext[1]{Here, all outcomes are essentially identical up to a phase difference between $\ket{00}$ and $\ket{11}$.}
\end{table}

In the following, we investigate the more general conditions under which a \emph{maximally} entangled state can be achieved with our protocol. With an explicit calculation performed the expressions of a reduced state can be readily determined. Remarkably, we find that there exists a condition under which the post-selected state $\widetilde{\rho}_{AB}^{-}$ is \emph{always} a maximally entangled state. This result holds true regardless of the specific time $t$ at which the measurement is performed, making the outcome particularly robust. Let us define the parameter $\mathcal{R}$ to be the ratio between the coefficients $c_{M,A}^X$ and $c_{N,B}^X$, thus $\mathcal{R}=c_{M,A}^X/c_{N,B}^X$. The condition for achieving maximally entangled state is that $\mathcal{R}^2=1$. In the cases where this condition is satisfied, the joint system behaves in the following specific way:
\begin{align}
	\label{eq:rhoabentries_ijkl}
	\bra{ij}^{AB}\widetilde{\rho}_{AB}^{-}\ket{kl}^{AB}=
    \begin{cases}
      \mathcal{G}/2 & \text{$(i,k=s$ and $j,l=1-s)$}\\
      -\mathcal{R}\mathcal{G}/2 & \text{$(i,l=s$ and $j,k=1-s)$}\\
      0 & \text{otherwise}.
    \end{cases}
\end{align}
Here $s=0$ or $1$, with $\mathcal{G}$ being a function of three parameters $c_{M,A}^X$, $\omega_z$, $t$, and its explicit expression is 
\begin{align}
	\label{eq:defg}\nonumber
	\mathcal{G}&=({c_{M,A}^X})^2\sin^2({\omega_z t/2})\sin({\Theta t})\\
	&\times\frac{(2\omega_z^2+{c_{M,A}^X}^2/2)\sin({\Theta t})+({c_{M,A}^X}^2/2)\sin{(3 \Theta t)}}{16\Theta^4},
\end{align}
where the frequency term is given by $\Theta=\sqrt{\omega_z^2+{c_{M,A}^X}^2}/2$ (see Appendix~\ref{sec:der_2021} for the derivation of the above two expressions). Employing the above results one can generate Bell states, for example, under the specific case where $\mathcal{R}=-1$, the resulting reduced state is the well-known maximally entangled Bell state $(\ket{01}^{AB}+\ket{10}^{AB})/\sqrt{2}$. The conditions for generating other prominent maximally entangled Bell states, including the initial state conditions and the auxiliary Hamiltonian forms, are summarized in Table~\ref{tab:table1}. 

Regarding physical implementation, our protocol is naturally suited for architectures where a control system coherently selects the order of local operations. A concrete photonic implementation can be based on three independent degrees of freedom (DoFs) of a single photon---using the path DoF as the control and two others (e.g., polarization and time-bin) as the target qubits, where each is addressed by localized optical elements for channels $\mathcal{M}$ and $\mathcal{N}$. Alternatively, one could envisage experimental implementations where the control and the two targets are mapped to distinct spatial modes that are routed through separate interferometric networks. Extending this architecture to matter-based platforms, such as trapped ions or superconducting circuits, would also be a worthwhile endeavor.

\section{Conclusion}
\label{sec:6}
In this work, our investigation into generating entanglement between non-interacting qubits using the quantum switch protocol led to the derivation of a fundamental trade-off relation. We then established that this inequality is a universal principle that governs a broad class of processes built from superpositions, not limited to those with ICO.

The central discovery of this work is that the structure of this universal bound predicts, and our concrete demonstrations confirm the existence of a ``quasi-deterministic'' mode of entanglement generation. We have shown that it is physically possible to find conditions where a maximally entangled state is guaranteed to be produced, that is all the measurement outcomes on an auxiliary control system lead to states with a maximal amount of entanglement. We have illustrated the rich variety of dynamical behaviors governed by our inequality, from trivial and complicated cases to the remarkable ``quasi-deterministic'' limit. Our results provide a novel and fundamental insight into the limits of entanglement generation via superpositions of non-interacting Hamiltonian-induced dynamics. We have also framed the quantum switch as a novel physical architecture for realizing this general principle.

While this study has focused on strictly interaction-free Hamiltonians, an interesting future direction is to investigate how introducing weak interactions would alter the entanglement generation protocol and the trade-off relation. Future works could also explore the fundamental impact of ICO on Hamiltonian dynamics, particularly within quantum many-body systems. Understanding this interplay would allow for the precise engineering of these dynamics, offering advanced control over the final entangled state and opening new avenues to optimize the resource efficiency of a protocol~\cite{PhysRevLett.122.120503, Adesso_2016, PhysRevLett.113.140401}. 

It is also of fundamental significance to draw insights from quantum multiplexing~\cite{PhysRevA.99.022337,PhysRevA.104.062409,PRXQuantum.3.040319,PhysRevA.107.022428, Sheng2023}. While our work focuses on the fundamental probabilistic limitation arising from post-selection in superpositions of non-interacting processes, it would be interesting in future work to explore how multiplexing strategies could be incorporated into our protocol to boost overall entanglement generation rates at the architecture level.

Furthermore, a heralded entangling gate between non-interacting systems can be viewed as our protocol applied to a chosen product input, with the heralding event playing the role of post-selection. For any fixed input state, the resulting output entanglement and success probability may therefore be constrained by a relation of the form similar to our central result. Obtaining a genuine, input-independent bound on the \emph{entangling power} of such gates would require optimizing this trade-off over all product inputs worths investigating~\cite{PhysRevA.62.030301}.

Finally, because our bound is derived only for unitary and non-interacting evolutions. However, for general multi-Kraus (i.e., more than two) CPTP maps, the dynamics is typically more complicated. Thus, the present trade-off should be understood as an ideal, unitary-limit result. Extending our analysis to noisy channels with higher-dimensional control systems is nontrivial, and constitutes a natural and meaningful direction for future work.

\emph{Note added.}--- During the preparation of this work, we became aware of a related paper~\cite{koudia2021causalactivation} by Koudia et al.. Our work is distinguished by the discovery of a fundamental trade-off relation and its generality for a broader class of coherent control schemes beyond ICO.

\section*{Data Availability}
The source code that supports the findings of this study are openly available at~\cite{entanglement_code}.
\begin{acknowledgements}
Y.C. acknowledges support by JSPS KAKENHI Grant Number 24KJ0718 and JST SPRING Grant Number JPMJSP2108. Y. H. acknowledges support by JSPS KAKENHI Grant Number JP22H03659, JP23K24915 and JP24K03008.
\end{acknowledgements}

\appendix
\section{Proof of the Concurrence-Probability Bound}
\label{sec:proof}

\subsection{Formulation of the Optimization Problem}
The following derivation for the trade-off relation is general. It assumes only the interference of two arbitrary and valid quantum evolution paths, resulting from any non-interacting unitaries acting on arbitrary separable initial states. The proof deliberately avoids imposing structural constraints between these paths such as the relationship specific to the ICO protocol. This construction is the key to the universality of the inequality, establishing it as a fundamental bound for this entire class of processes. The specific physical models explored in the main text are then presented as concrete examples to test whether this universal bound can be saturated by a physically realizable system.

Therefore, the primary objective of this section is to provide a self-contained derivation of the trade-off relation presented in Eq.~\eqref{eq:pc_trade_off}. To achieve this, we will determine the maximum achievable concurrence, $\mathcal{C}$, for any given success probability, $P_\text{succ}$. Our methodology involves treating $P_\text{succ}$ as a fixed, but arbitrary parameter that defines the region for our optimization problem. The final result will be an inequality that represents the upper bound on concurrence as a function of this parameter, i.e., $\mathcal{C}_\text{max}=f(P_\text{succ})$. This function delineates the boundary of what is physically possible with our protocol.

We begin by considering the post-selected two-qubit pure state, $\ket{\widetilde{\Psi}}^{AB} = \sum_{k,l}p_{kl}\ket{kl}^{AB}$, where $\{p_{kl}\}$ are the complex amplitudes resulting from the dynamics of quantum superpositions of process and subsequent post-selection. The concurrence of this pure state can be calculated from the purity of its reduced states. The reduced state of qubit $\mathcal{Q}^A$ is obtained by performing a partial trace over the degrees of freedom of qubit $\mathcal{Q}^B$ , i.e., $\widetilde{\rho}_{A}=\text{tr}_B\ketbra{\widetilde{\Psi}}^{AB}$.

\subsubsection{Derivation of the Reduced Density Matrix}
To make this proof self-contained, we now explicitly derive the matrix form of $\widetilde{\rho}_{A}$. The state vector can be written in the computational basis as:
\begin{align}\nonumber
    \ket{\widetilde{\Psi}}^{AB} &= p_{00}\ket{00}^{AB} + p_{01}\ket{01}^{AB} \\
    &~~~+ p_{10}\ket{10}^{AB} + p_{11}\ket{11}^{AB}.
\end{align}
The density operator for the joint system is $\widetilde{\rho}_{AB} = \ketbra{\widetilde{\Psi}}^{AB}$. The elements of the reduced density matrix for $\mathcal{Q}^A$, that is $\widetilde{\rho}_{A} = \text{tr}_B \ketbra{\widetilde{\Psi}}^{AB}$ are given by the evaluation $(\widetilde{\rho}_A)_{ij} = \sum_{k \in \{0,1\}} \bra{i,k}^{AB} \widetilde{\rho}_{AB} \ket{j,k}^{AB}$.

Let us compute the diagonal elements first. For the $(0,0)$ element, we have
\begin{align}\nonumber
    (\widetilde{\rho}_A)_{00} &= \sum_{k \in \{0,1\}} \bra{0,k}^{AB} \left( \ket{\widetilde{\Psi}}^{AB}\bra{\widetilde{\Psi}}^{AB} \right) \ket{0,k}^{AB} \\\nonumber
    &= \braket{00}{\widetilde{\Psi}}^{AB}\braket{\widetilde{\Psi}}{00}^{AB} + \braket{01}{\widetilde{\Psi}}^{AB}\braket{\widetilde{\Psi}}{01}^{AB} \\
    &= |p_{00}|^2 + |p_{01}|^2.
\end{align}
For the $(1,1)$ element, we perform a similar calculation,
\begin{align}\nonumber
    (\widetilde{\rho}_A)_{11} &= \sum_{k \in \{0,1\}} \bra{1,k}^{AB} \left( \ket{\widetilde{\Psi}}^{AB}\bra{\widetilde{\Psi}}^{AB} \right) \ket{1,k}^{AB} \\\nonumber
    &= \braket{10}{\widetilde{\Psi}}^{AB}\braket{\widetilde{\Psi}}{10}^{AB} + \braket{11}{\widetilde{\Psi}}^{AB}\braket{\widetilde{\Psi}}{11}^{AB} \\
    &= |p_{10}|^2 + |p_{11}|^2.
\end{align}
Next, we compute the off-diagonal elements. For the $(1,0)$ element, we have
\begin{align}\nonumber
    (\widetilde{\rho}_A)_{10} &= \sum_{k \in \{0,1\}} \bra{1,k}^{AB} \left( \ket{\widetilde{\Psi}}^{AB}\bra{\widetilde{\Psi}}^{AB} \right) \ket{0,k}^{AB} \\\nonumber
    &= \braket{10}{\widetilde{\Psi}}^{AB}\braket{\widetilde{\Psi}}{00}^{AB} + \braket{11}{\widetilde{\Psi}}^{AB}\braket{\widetilde{\Psi}}{01}^{AB} \\
    &= p_{10}\bar{p}_{00} + p_{11}\bar{p}_{01}.
\end{align}
Finally, since a density matrix must be Hermitian, the $(0,1)$ element is the complex conjugate of the $(1,0)$ element:
\begin{align}\nonumber
    (\widetilde{\rho}_A)_{01} &= ((\widetilde{\rho}_A)_{10})^* \\\nonumber
    &= (p_{10}\bar{p}_{00} + p_{11}\bar{p}_{01})^* \\
    &= \bar{p}_{10}p_{00} + \bar{p}_{11}p_{01}.
\end{align}
We can now express these matrix elements using the quantities $P_\text{succ}$, $Q_1$, and $Q_2$. These are defined as follows,
\begin{align}\nonumber
    Q_1 &\coloneqq \sum_{k,l}(-1)^{k} |p_{k,l}|^2 \\
    &= |p_{00}|^2 + |p_{01}|^2 - |p_{10}|^2 - |p_{11}|^2, \\\nonumber
    Q_2 &\coloneqq \sum_{k}p_{0,k}\bar{p}_{1,k} \\
    &= p_{00}\bar{p}_{10} + p_{01}\bar{p}_{11}, \\
    P_\text{succ} &\coloneqq \sum_{k,l}\abs{p_{kl}}^2.
\end{align}
Using these definitions, we can see that $(\widetilde{\rho}_A)_{00} = \frac{1}{2}(P_\text{succ} + Q_1)$ and $(\widetilde{\rho}_A)_{11} = \frac{1}{2}(P_\text{succ} - Q_1)$. For the off-diagonal elements, we see that $(\widetilde{\rho}_A)_{01} = Q_2$ and $(\widetilde{\rho}_A)_{10} = \overline{Q_2}$. Assembling these components gives the reduced density matrix:
\begin{align}
\label{eq:rho_a_tilde}
	\widetilde{\rho}_{A} =					
    \begin{bmatrix}
	    \frac{1}{2}(P_\text{succ}+Q_1) & Q_2 \\
        \overline{Q_2}  & \frac{1}{2}(P_\text{succ}-Q_1) 
    \end{bmatrix}.
\end{align}

\subsubsection{Optimization Strategy}
From Eq.~\eqref{eq:C_def_rhoa}, we calculate the concurrence of the final state.
Recall that,
\begin{align}
	\mathcal{C}\bigg(\ket{\widetilde{\Psi}}^{AB}\bigg) = \sqrt{2\left[1 - \frac{\text{Tr}(\tilde{\rho}_A^2)}{(\text{Tr}(\tilde{\rho}_A))^2}\right]}.
\end{align}
Here, the denominator $(\text{Tr}(\tilde{\rho}_A))^2$ is indeed $P_{\text{succ}}^2$. Based on Eq.~\eqref{eq:rho_a_tilde}, we can calculate the trace of the matrix-squared:
\begin{align}\nonumber
	\text{Tr}(\tilde{\rho}_A^2) &= \text{Tr}\left(\begin{bmatrix} \frac{1}{2}(P_{\text{succ}}+Q_{1}) & Q_{2} \\ \bar{Q}_{2} & \frac{1}{2}(P_{\text{succ}}-Q_{1}) \end{bmatrix}^2\right) \\
	&= \frac{1}{2}(P_{\text{succ}}^2 + Q_1^2) + 2|Q_2|^2.	
\end{align}
        
Substituting this into Eq.~\eqref{eq:C_def_rhoa} for $\mathcal{C}^2$ gives:
\begin{align}\nonumber
        \mathcal{C}^2\bigg(\ket{\widetilde{\Psi}}^{AB}\bigg) &= 2\left[1 - \frac{\text{Tr}(\tilde{\rho}_A^2)}{(\text{Tr}(\tilde{\rho}_A))^2}\right] \\\nonumber
        &= 2\left[1 - \frac{\frac{1}{2}(P_{\text{succ}}^2 + Q_1^2) + 2|Q_2|^2}{P_{\text{succ}}^2}\right] \\\nonumber
      &= 2 \left[ \frac{P_{\text{succ}}^2 - \frac{1}{2}P_{\text{succ}}^2 - \frac{1}{2}Q_1^2 - 2|Q_2|^2}{P_{\text{succ}}^2} \right] \\\nonumber
      &= \frac{P_{\text{succ}}^2 - Q_1^2 - 4|Q_2|^2}{P_{\text{succ}}^2} \\
   &= 1 - \frac{Q_1^2 + 4|Q_2|^2}{P_{\text{succ}}^2}.
\end{align}

After taking the square root, we arrive at:
\begin{equation}
    \mathcal{C}\bigg(\ket{\widetilde{\Psi}}^{AB}\bigg)=\sqrt{1-\frac{Q_{1}^{2}+4|Q_{2}|^{2}}{P_{\text{succ}}^{2}}}.
    \label{eq:concurrence_explicit}
\end{equation}
Our goal is to maximize $\mathcal{C}$ for a fixed $P_{\text{succ}}$. Maximizing a function of the form $\sqrt{1-x}$ is equivalent to minimizing the argument $x$. Therefore, our task is equivalent to minimizing the term $(Q_{1}^{2}+4|Q_{2}|^{2})/P_{\text{succ}}^{2}$. Since we are treating $P_{\text{succ}}$ as a constant parameter in this optimization, our task simplifies to finding the minimum possible value of the numerator, $Q_{1}^{2}+4|Q_{2}|^{2}$, subject to the constraint that the total probability is fixed, i.e., $\sum_{kl}|p_{kl}|^2 = P_{\text{succ}}$.

To facilitate this minimization, it is useful to introduce an auxiliary quantity, $\mathcal{D}$. We define this quantity as
\begin{align}
	\label{eq:D_def}
    \mathcal{D} &:= Q_{1}^{2}+4|Q_{2}|^{2}-P_{\text{succ}}^{2}.
\end{align}
The relationship between the concurrence and $\mathcal{D}$ can be seen by rearranging this definition: $Q_{1}^{2}+4|Q_{2}|^{2} = \mathcal{D} + P_{\text{succ}}^{2}$. Substituting this into the expression for the squared concurrence gives
\begin{align}\nonumber
    \mathcal{C}^2 &= 1-\frac{\mathcal{D} + P_{\text{succ}}^{2}}{P_{\text{succ}}^{2}} \\\nonumber
    &= \frac{P_{\text{succ}}^{2} - (\mathcal{D} + P_{\text{succ}}^{2})}{P_{\text{succ}}^{2}} \\
    &= -\frac{\mathcal{D}}{P_\text{succ}^2}.
\end{align}
This shows that maximizing $\mathcal{C}$ is equivalent to finding the minimum of $\mathcal{D}$. The term $-P_{\text{succ}}^2$ in the definition of $\mathcal{D}$ is a constant offset and does not alter the location of the minimum, here it is included to simplify later expressions. The trivial maximum value of $\mathcal{D}$ is zero, which corresponds to a separable final state ($\mathcal{C}=0$), which is the reason why we focus on finding its nontrivial minimum.

\subsection{Representing the Constraints}
We use the general operators $\mathbb{U}$ and $\mathbb{V}$ to emphasize the universality of our result, as any physical process involving the superposition of two distinct, non-interacting evolution paths can be mapped to this abstract form (e.g., by setting $\mathbb{U} = U_\mathcal{M} U_\mathcal{N}$ and $\mathbb{V} = U_\mathcal{N} U_\mathcal{M}$), ensuring our derived inequality is broadly applicable.

To find the minimum of $\mathcal{D}$, we start with analyzing the constraints imposed on the amplitudes $\{p_{kl}\}$ by the physical structure of the protocol. From the form of Eq.~\eqref{eq:deltaUrho}, one can see that the post-selected state is given by the expression $\ket{\widetilde{\Psi}}^{AB} = \frac{1}{2}(\mathbb{U}-\mathbb{V})\ket{\Psi}^{AB}$. A crucial feature of our protocol is that the initial state $\ket{\Psi}^{AB}$ is separable, and the operators $\mathbb{U}$ and $\mathbb{V}$ are local, meaning they cannot create entanglement by themselves. Consequently, the states $\mathbb{U}\ket{\Psi}^{AB}$ and $\mathbb{V}\ket{\Psi}^{AB}$ must also be separable.

\subsubsection{Revisiting the Separability Condition for Pure Bipartite States}
Before proceeding, it is instructive to establish the mathematical condition for the separability of a generic two-qubit pure state, $\ket{\psi}^{AB}$. By definition, such a state is separable if and only if it can be written as a tensor product of two single-qubit states, $\ket{\phi}^A$ and $\ket{\chi}^B$. Let us write these single-qubit states in their general form:
\begin{align}
    \ket{\phi}^A &= \alpha\ket{0}^A + \beta\ket{1}^A, \\
    \ket{\chi}^B &= \gamma\ket{0}^B + \delta\ket{1}^B,
\end{align}
where $\alpha, \beta, \gamma, \delta$ are complex numbers. The tensor product state is then
\begin{align}\nonumber
    \ket{\psi}^{AB} &= \ket{\phi}^A \otimes \ket{\chi}^B \\\nonumber
    &= (\alpha\ket{0}^A + \beta\ket{1}^A) \otimes (\gamma\ket{0}^B + \delta\ket{1}^B) \\\nonumber
    &= \alpha\gamma\ket{00}^{AB} + \alpha\delta\ket{01}^{AB}\\
     &~~~+ \beta\gamma\ket{10}^{AB} + \beta\delta\ket{11}^{AB}.
\end{align}
Let us compare this to the general expression for a two-qubit state written with amplitudes $\{x_{kl}\}$,
\begin{align}\nonumber
    \ket{\psi}^{AB} &= x_{00}\ket{00}^{AB} + x_{01}\ket{01}^{AB} \\
         &~~~+ x_{10}\ket{10}^{AB} + x_{11}\ket{11}^{AB}.
\end{align}
By comparing the two expressions, we can identify the coefficients for a two-qubit state,
\begin{align}
    x_{00} = \alpha\gamma, \quad x_{01} = \alpha\delta, \quad x_{10} = \beta\gamma, \quad x_{11} = \beta\delta.
\end{align}
We now test the relationship between these coefficients. Let us compute the product of the diagonal coefficients and the product of the off-diagonal coefficients,
\begin{align}\nonumber
    x_{00}x_{11} &= (\alpha\gamma)(\beta\delta) \\
    &= \alpha\beta\gamma\delta, \\\nonumber
    x_{01}x_{10} &= (\alpha\delta)(\beta\gamma) \\
    &= \alpha\beta\gamma\delta.
\end{align}
For any separable state, these two products must be equal. This gives the following condition for the separability of a pure two-qubit state:
\begin{align}
    x_{00}x_{11} - x_{01}x_{10} = 0.
\end{align}
This condition is precisely equivalent to stating that the determinant of the coefficient matrix of the state, when arranged as $M = \begin{bsmallmatrix} x_{00} & x_{01} \\ x_{10} & x_{11} \end{bsmallmatrix}$, must be zero. This facilitates representing the constraints we will develop in the following.

\subsubsection{Separability Constraints in the Vector Representation}
\label{sec:pure_cond}
To manage the constraints systematically, we represent the two separable states, $\frac{1}{2}\mathbb{U}\ket{\Psi}^{AB}$ and $\frac{1}{2}\mathbb{V}\ket{\Psi}^{AB}$, as vectors in an eight-dimensional real vector space. Let the complex amplitudes of these two intermediate states be denoted by $\{q_{kl}\}$ and $\{s_{kl}\}$ respectively. We can decompose them into real and imaginary parts,
\begin{align}\nonumber
   ~ q_{00} = \frac{1}{2}(A_1+\mathbf{i}E_1), \quad q_{01} = \frac{1}{2}(B_1+\mathbf{i}F_1),\\ 
        ~q_{10} = \frac{1}{2}(C_1+\mathbf{i}G_1), \quad q_{11} = \frac{1}{2}(D_1+\mathbf{i}H_1).
\end{align}
This allows us to define two real vectors, $V_1$ and $V_2$, as follows
\begin{align}
    V_1 &= (A_1, B_1, C_1, D_1, E_1, F_1, G_1, H_1), \\
    V_2 &= (A_2, B_2, C_2, D_2, E_2, F_2, G_2, H_2).
\end{align}
The physical properties of the states impose several constraints on these vectors. First, the normalization of the states requires that the sum of the squared magnitudes of their amplitudes is unity. This leads to our first constraint on the norms of the vectors,
\begin{align}\nonumber
    \textbf{C}_1:~~~ \|V_i\|^2  = A_i^2+B_i^2+\dots+H_i^2 = 1, \\
    \quad \text{for } i \in \{1,2\}.~~~~~~~~~~~~~
\end{align}
Second, applying the separability condition derived above to the intermediate states gives the origin of constraints $\textbf{C}_2$ and $\textbf{C}_3$. For the state represented by the vector $V_i$, as we have discussed earlier in Sec.~\ref{sec:pure_cond}, one can write the separability condition as:
\begin{align}\nonumber
    &\frac{1}{4}(A_i+\mathbf{i}E_i)(D_i+\mathbf{i}H_i) \\
    - &\frac{1}{4}(B_i+\mathbf{i}F_i)(C_i+\mathbf{i}G_i) = 0.
\end{align}
Furthermore, by expanding the products, we get that
\begin{align}\nonumber
\label{eq:cond23}
    &(A_iD_i + \mathbf{i}A_iH_i + \mathbf{i}E_iD_i - E_iH_i) \\
    &- (B_iC_i + \mathbf{i}B_iG_i + \mathbf{i}F_iC_i - F_iG_i) = 0.
\end{align}
In the above expressions, the dummy index $i$ should not be confused with the imaginary unit $\mathbf{i}$. Therefore, for this complex number to be zero, both its real and imaginary parts must be zero. Let us group the real-part terms,
\begin{align}\nonumber
    \text{Re}[\text{LHS of Eq.}~\eqref{eq:cond23}] &= A_iD_i - E_iH_i - B_iC_i + F_iG_i\\
     &= 0.
\end{align}
Therefore it gives our second constraint,
\begin{align}
    \textbf{C}_2:~~~ A_iD_i - E_iH_i - B_iC_i + F_iG_i &= 0.
\end{align}
Next, let us group the imaginary-part terms,
\begin{align}\nonumber
    \text{Im}[\text{LHS of Eq.}~\eqref{eq:cond23}] &= A_iH_i + E_iD_i - B_iG_i - F_iC_i\\
     &= 0.
\end{align}
Furthermore, we have the following third constraint,
\begin{align}
    \textbf{C}_3:~~~ A_iH_i + E_iD_i - B_iG_i - F_iC_i &= 0.
\end{align}
Finally, the amplitudes $\{p_{kl}\}$ of our final state $\ket{\widetilde{\Psi}}^{AB}$ are given by the difference $p_{kl} = q_{kl} - s_{kl}$. This means the vector of real and imaginary components of amplitudes of the final state is given by the difference vector $\Delta V = V_1 - V_2$. The success probability $P_\text{succ}$ is therefore related to the squared norm of this difference vector:
\begin{align}\nonumber
    \textbf{C}_4:~~~ \lVert \Delta V \rVert^2 &= \sum (\Delta A)^2 + \dots + (\Delta H)^2 \\
    &= 4P_\text{succ},
\end{align}
where 
\begin{align}
\Delta A &= A_1 - A_2,~~\dots,~~\Delta H = H_1 - H_2.
\end{align}With this detailed conditions established, we can proceed further to find the minimum of $\mathcal{D}$.

\subsection{Derivation of the Central Identity}

We now show that the quantity $\mathcal{D}$ can be expressed in a much more compact form. A straightforward but lengthy algebraic expansion of Eq.~\eqref{eq:D_def} reveals a direct connection to the determinant of the coefficient matrix. We begin by expanding the term $Q_1^2 - P_\text{succ}^2$:
\begin{align}
    Q_1^2 - P_\text{succ}^2 = (Q_1 - P_\text{succ})(Q_1 + P_\text{succ}).
\end{align}
Substituting the definitions of $Q_1$ and $P_\text{succ}$ yields
\begin{align}\nonumber
    Q_1 - P_\text{succ} &= -2(|p_{10}|^2 + |p_{11}|^2) \\
    &= -2\sum_{l}|p_{1l}|^2, \\\nonumber
    Q_1 + P_\text{succ} &= 2(|p_{00}|^2 + |p_{01}|^2) \\
    &= 2\sum_{k}|p_{0k}|^2.
\end{align}
Thus, we have the intermediate result,
\begin{align}
    Q_1^2 - P_\text{succ}^2 = -4\left(\sum_{k}|p_{0k}|^2\right)\left(\sum_{l}|p_{1l}|^2\right).
\end{align}
Now, substituting this back into the definition of $\mathcal{D}$:
\begin{align}\nonumber
    \mathcal{D} &= -4\left(|p_{00}|^2+|p_{01}|^2\right)\left(|p_{10}|^2+|p_{11}|^2\right)  \\\nonumber
    & \qquad + 4|p_{00}\bar{p}_{10} + p_{01}\bar{p}_{11}|^2 \\
    &= -4\Big( |p_{00}|^2|p_{10}|^2 + |p_{00}|^2|p_{11}|^2 \nonumber \\
    & \qquad \quad + |p_{01}|^2|p_{10}|^2 + |p_{01}|^2|p_{11}|^2 \Big) \nonumber \\
    & \quad + 4\Big( |p_{00}|^2|p_{10}|^2 + p_{00}\bar{p}_{10}\bar{p}_{01}p_{11} \nonumber \\
    & \qquad \quad + \bar{p}_{00}p_{10}p_{01}\bar{p}_{11} + |p_{01}|^2|p_{11}|^2 \Big).
\end{align}
Then we are left with:
\begin{align}\nonumber
    \mathcal{D} &= -4|p_{00}|^2|p_{11}|^2 - 4|p_{01}|^2|p_{10}|^2  \\\nonumber
    & \quad + 4p_{00}\bar{p}_{10}\bar{p}_{01}p_{11} + 4\bar{p}_{00}p_{10}p_{01}\bar{p}_{11} \\\nonumber
    &= -4 \Big( |p_{00}|^2|p_{11}|^2 + |p_{01}|^2|p_{10}|^2  \\ \nonumber
    & \qquad \quad - p_{00}\bar{p}_{10}\bar{p}_{01}p_{11} - \bar{p}_{00}p_{10}p_{01}\bar{p}_{11} \Big) \\\nonumber
    &= -4 \Big( |p_{00}p_{11}|^2 + |p_{01}p_{10}|^2  \\
    & \qquad \quad - \left( (p_{00}p_{11})(\overline{p_{01}p_{10}}) + (\overline{p_{00}p_{11}})(p_{01}p_{10}) \right) \Big)
\end{align}
By a quick observation, one notices that the above expression is in fact the expansion of the following compact form,
\begin{align}
    \mathcal{D} &= -4|p_{00}p_{11} - p_{01}p_{10}|^2.
    \label{eq:D_expanded}
\end{align}
Furthermore, the term $|p_{00}p_{11} - p_{01}p_{10}|^2$ is precisely the squared magnitude of the determinant of the matrix of coefficients $\{p_{kl}\}$. To relate this back to our previous conditions, we now formally introduce the matrix representation, $M_i$, for each vector $V_i$:
\begin{align}
    M_i &=
    \begin{bmatrix}
        A_i + iE_i & B_i + iF_i \\
        C_i + iG_i & D_i + iH_i
    \end{bmatrix}.
\end{align}
The constraints $\textbf{C}_2$ and $\textbf{C}_3$ are now compactly expressed as 
\begin{align}
\label{eq:detzero}
	\det M_i = 0.
\end{align}

 Next, let us define the quantity $\mathcal{I} = |\det(M_1 - M_2)|^2$. Since the coefficients $\{p_{kl}\}$ are the elements of $\frac{1}{2}(M_1 - M_2)$, we have:
\begin{align}\nonumber
    |p_{00}p_{11} - p_{01}p_{10}|^2 &= \left|\det\left(\frac{1}{2}(M_1-M_2)\right)\right|^2 \\\nonumber
    &= \frac{1}{16}|\det(M_1-M_2)|^2 \\
    &= \frac{\mathcal{I}}{16}.
\end{align}
Substituting this into Eq.~\eqref{eq:D_expanded} yields the simple and direct identity,
\begin{align}\nonumber
    \mathcal{D} &= -4 \left(\frac{\mathcal{I}}{16}\right) \\
    &= -\frac{\mathcal{I}}{4}.
    \label{eq:D_I_relation}
\end{align}
This relation provides the crucial link between our optimization quantity $\mathcal{D}$ and the auxiliary quantity $\mathcal{I}$. It means that finding the minimum of $\mathcal{D}$ is equivalent to finding the maximum of $\mathcal{I}$.

\subsection{Bounding the Determinant via Duality}
We now proceed to find the upper bound on $\mathcal{I}$. 
First, let us $u$ and $v$ be the column vectors of $ M_1-M_2 $. The magnitude of the determinant satisfies:
\begin{align}
\lvert \det (M_1-M_2) \rvert \leq \|u\| \cdot \|v\|.
\end{align}
On the other hand, the Frobenius norm squared of $ M_1-M_2 $ is
\begin{align}\nonumber
\|M_1-M_2\|_F^2 = &\|u\|^2 + \|v\|^2\\
=&(\Delta A)^2 + (\Delta B)^2 + \dots + (\Delta H)^2,
\end{align}
which is exactly $4P_\text{succ}$. Therefore, we have the following inequality,
\begin{align}
\|u\| \cdot \|v\| \leq \frac{\|u\|^2 + \|v\|^2}{2} = \frac{\|M_1-M_2\|_F^2}{2}.
\end{align}
Combining these results, we obtain
\begin{align}
\lvert\det (M_1 - M_2) \rvert \leq \frac{\|(M_1 - M_2)\|_F^2}{2}.
\end{align}

Applying the above result to the matrix $(M_1 - M_2)$ yields a bound on $\mathcal{I}$,
\begin{align}\nonumber
    \mathcal{I} &\le \frac{1}{4}\|M_1-M_2\|_F^4 \\\nonumber
    &= \frac{1}{4}\|\Delta V\|^4 \\\nonumber
    &= \frac{1}{4}(4P_\text{succ})^2 \\
    &= 4P_\text{succ}^2.
\end{align}
This bound, however, is not tight over the full domain $P_{\text{succ}} \in [0,1]$. To establish a tighter bound, we use a fundamental symmetry of the protocol. The post-selection measurement on the control system has two orthogonal outcomes. Our analysis has focused on the $|-\rangle^C$ outcome, yielding probability $P_{\text{succ}}$. The orthogonal $|+\rangle^C$ outcome yields the state $\ket{\widetilde{\Psi}'}^{AB} = \frac{1}{2}(\mathbb{U}+\mathbb{V})\ket{\Psi}^{AB}$ with probability $P'_{\text{succ}} = 1 - P_{\text{succ}}$.

The mathematical structure for this dual problem is identical. The quantity analogous to $\mathcal{I}$ for this outcome is $\mathcal{J} = |\det(M_1+M_2)|^2$. Applying the same logic, $\mathcal{J}$ is bounded by $4(P'_{\text{succ}})^2 = 4(1-P_{\text{succ}})^2$. A key property relating the two determinants is that $\det(M_1-M_2) = -\det(M_1+M_2)$, which follows from the condition that $\det M_1 = \det M_2 = 0$. This implies $\mathcal{I} = \mathcal{J}$, and therefore $\mathcal{I}$ must also satisfy the bound from the dual problem, $\mathcal{I} \le 4(1-P_{\text{succ}})^2$.

\subsubsection{Proof of the Determinant Identity}
The assertion that $\det(M_1-M_2) = -\det(M_1+M_2)$ given $\det M_1 = \det M_2 = 0$ is a crucial step that warrants a formal proof. Let $M_1$ and $M_2$ be arbitrary $2 \times 2$ complex matrices:
\begin{align}
    M_1 = \begin{bmatrix} a & b \\ c & d \end{bmatrix}, \quad M_2 = \begin{bmatrix} e & f \\ g & h \end{bmatrix}.
\end{align}
The condition in Eq.~\eqref{eq:detzero}, that is $\det M_1 = 0$ implies $ad-bc=0$. Similarly, $\det M_2 = 0$ implies $eh-fg=0$.
Let us compute the determinant of the sum, that is for $\det(M_1+M_2)$ we have
\begin{align}\nonumber
    \det(M_1+M_2) &= \det \begin{bmatrix} a+e & b+f \\ c+g & d+h \end{bmatrix} \\\nonumber
    &= (a+e)(d+h) - (b+f)(c+g) \\\nonumber
    &= (ad+ah+ed+eh) \nonumber \\\nonumber
    & \qquad - (bc+bg+fc+fg) \\
    &= (ad-bc) + (eh-fg) \nonumber \\
    & \qquad + ah+ed-bg-fc.
\end{align}
Since $\det M_1 = ad-bc = 0$ and $\det M_2 = eh-fg = 0$, this simplifies to
\begin{align}
    \det(M_1+M_2) = ah+ed-bg-fc.
\end{align}
Next, we compute the determinant of the difference, $\det(M_1-M_2)$:
\begin{align}\nonumber
    \det(M_1-M_2) &= \det \begin{bmatrix} a-e & b-f \\ c-g & d-h \end{bmatrix} \\\nonumber
    &= (a-e)(d-h) - (b-f)(c-g) \\\nonumber
    &= (ad-ah-ed+eh) \nonumber \\\nonumber
    & \qquad - (bc-bg-fc+fg) \\
    &= (ad-bc) + (eh-fg) \nonumber \\
    & \qquad - ah-ed+bg+fc.
\end{align}
Again, using the fact that the determinants of $M_1$ and $M_2$ are zero, this simplifies to
\begin{align}
    \det(M_1-M_2) = -ah-ed+bg+fc.
\end{align}
By comparing the two final expressions, we can see the desired relationship
\begin{align}\nonumber
    \det(M_1-M_2) &= - (ah+ed-bg-fc) \\
    &= -\det(M_1+M_2).
\end{align}
This completes the proof of the identity. This result means that $|\det(M_1-M_2)|^2 = |\det(M_1+M_2)|^2$, and therefore $\mathcal{I} = \mathcal{J}$.

\subsection{Finalizing the Bound}
Since $\mathcal{I} = \mathcal{J}$, the bound derived for the dual problem must also apply to $\mathcal{I}$. Therefore, both inequalities for $\mathcal{I}$ must hold simultaneously:
\begin{align}
    \mathcal{I} \le 4P_{\text{succ}}^{2} \quad \text{and} \quad \mathcal{I} \le 4(1-P_{\text{succ}})^{2}.
\end{align}
This implies that $\mathcal{I}$ must be less than or equal to the minimum of the two bounds, giving us the tight upper bound
\begin{equation}
    \mathcal{I} \le \min\{4P_{\text{succ}}^{2}, 4(1-P_{\text{succ}})^{2}\}.
\end{equation}
Finally, we substitute this tight bound back into our central identity from Eq.~\eqref{eq:D_I_relation}. Reversing the inequality sign we obtain the lower bound on $\mathcal{D}$, that is
\begin{align}\nonumber
    \mathcal{D} &\ge -\frac{1}{4}\min\{4P_{\text{succ}}^{2}, 4(1-P_{\text{succ}})^{2}\} \\
    &= -\min\{P_{\text{succ}}^{2}, (1-P_{\text{succ}})^{2}\}.
\end{align}
This result then servers to directly derive the trade-off relation for concurrence.

To make the final result explicit, we now derive the trade-off relation for the concurrence itself. We have established that $\mathcal{C}^2 = -\mathcal{D}/P_\text{succ}^2$, which implies that the maximum possible squared concurrence is $\mathcal{C}_\text{max}^2 = -\mathcal{D}_\text{min}/P_\text{succ}^2$. We consider two cases based on our bound for $\mathcal{D}_\text{min}$.

\textbf{Case 1:} $0 \le P_\text{succ} \le 1/2$. In this regime, $P_\text{succ}^2 \le (1-P_\text{succ})^2$, so $\mathcal{D}_\text{min} = -P_\text{succ}^2$. The maximum concurrence is
\begin{align}\nonumber
    \mathcal{C}_\text{max}^2 &= -\frac{-P_\text{succ}^2}{P_\text{succ}^2} \\
    &= 1.
\end{align}

This implies that for any success probability up to $1/2$, it is possible to achieve maximal entanglement, $\mathcal{C}=1$.

\textbf{Case 2:} $1/2 \le P_\text{succ} \le 1$. In this regime, $(1-P_\text{succ})^2 \le P_\text{succ}^2$, so $\mathcal{D}_\text{min} = -(1-P_\text{succ})^2$. The maximum concurrence is:
\begin{align}\nonumber
    \mathcal{C}_\text{max}^2 &= -\frac{-(1-P_\text{succ})^2}{P_\text{succ}^2} \\
    &= \left(\frac{1-P_\text{succ}}{P_\text{succ}}\right)^2.
\end{align}
Taking the square root, we find $\mathcal{C}_\text{max} = (1-P_\text{succ})/P_\text{succ}$.
\begin{figure*}[t]
  \includegraphics[width=1.0\textwidth]{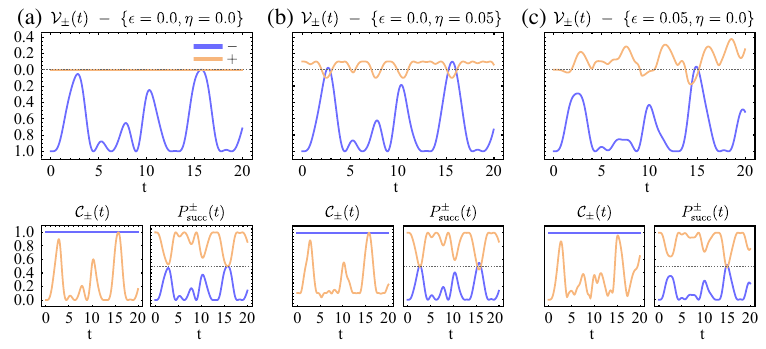}
  \centering
  \caption{Time evolution of the trade-off`violation $\mathcal{V}_\pm(t)$, concurrence $C_\pm(t)$, and success probability $P_{\mathrm{succ},\pm}(t)$ for three representative cases. The top row shows $\mathcal{V}_\pm(t) = P_{\mathrm{succ},\pm}(t)\bigl(1 + C_\pm(t)\bigr) - 1$ for (a) the baseline protocol with strictly non-interacting dynamics and separable input ($\epsilon = 0$, $\eta = 0$), (b) weak initial entanglement at fixed non-interacting dynamics ($\epsilon = 0$, $\eta = 0.05$), and (c) weak cross-talk at fixed separable input ($\epsilon = 0.05$, $\eta = 0$). Blue and yellow curves correspond to the $|-\rangle^C$ and $|+\rangle^C$ branches, respectively. The bottom rows show the corresponding concurrence $C_\pm(t)$ and success probability $P_{\mathrm{succ},\pm}(t)$ for each branch. In panel~(a) the inequality $P_{\mathrm{succ},\pm}(t)\bigl(1 + C_\pm(t)\bigr)\le 1$ is always satisfied, i.e., $\mathcal{V}_\pm(t)\le 0$. In panels~(b) and~(c) small positive excursions of $\mathcal{V}_+(t)$ appear, indicating that weak initial entanglement or weak interactions can induce slight violations of the interaction–free trade-off.}
  \label{fig:weak_1}
\end{figure*}
Combining these two cases gives the complete trade-off relation for the maximum achievable concurrence:
\begin{align}
	\mathcal{C}(\widetilde{\Psi}_{AB}) \le
	\begin{cases}
		1, &\text{if }0 \le P_{\text{succ}} \le \frac{1}{2}\\
		\frac{1-P_{\text{succ}}}{P_{\text{succ}}}, &\text{if }\frac{1}{2} \le P_{\text{succ}} \le 1.
	\end{cases}
\end{align}
This concludes the proof.

\section{Extension for Mixed Initial States}
\label{sec:mix_extension}
Let the two target systems be qubits. In each branch the target evolution is strictly local, and post-selection is equivalent to the effect induced by the operators of the form $M = \frac{1}{2}(\mathbb{U}-\mathbb{V}) \quad \text{and} \quad P = \frac{1}{2}(\mathbb{U}+\mathbb{V})$, where $\mathbb{U}$ and $\mathbb{V}$ are local unitaries.
For any separable mixed input state $\rho$ on qubits $AB$, the success probability $P_{\mathrm{succ}}(\rho)=\operatorname{Tr}\!\bigl[K\rho K^\dagger\bigr]$ and the concurrence $C\!\bigl(\rho_{\mathrm{out}}(\rho)\bigr)$ of the normalized postselected state $\rho_{\mathrm{out}}(\rho)=K\rho K^\dagger/P_{\mathrm{succ}}(\rho)$ satisfy, where $K=M$ or $P$.

\begin{align}
P_{\mathrm{succ}}(\rho)\,\bigl[1+C\bigl(\rho_{\mathrm{out}}(\rho)\bigr)\bigr]\;\le\;1.
\end{align}

\textit{Proof.--}

Since $\rho$ is separable, we express it as a convex combination of pure product states:
\begin{align}
  	  \rho=\sum_j p_j\,|\phi_j\rangle\!\langle\phi_j|,\qquad |\phi_j\rangle=|a_j\rangle\otimes|b_j\rangle,
\end{align}
where $\quad p_j\ge 0,\ \sum_j p_j=1$. Next, define $|\psi_j\rangle:=K|\phi_j\rangle$ and $s_j:=\langle\psi_j|\psi_j\rangle=\|K|\phi_j\rangle\|^2$. By linearity,
\begin{align}
  	P_{\mathrm{succ}}(\rho)=\operatorname{Tr}\!\bigl[K\rho K^\dagger\bigr]=\sum_j p_j s_j=S.
\end{align}

For $s_j>0$, let $|\varphi_j\rangle:=|\psi_j\rangle/\sqrt{s_j}$ and $C_j:=C(|\varphi_j\rangle)$.
  Under the stated assumptions (local branches, product input), our main result for pure-state applies to each $j$:

  \begin{align}
  	  s_j\,\bigl(1+C_j\bigr)\;\le\;1.
  \end{align}
The normalized postselected state is then:
  \begin{align}\nonumber
  \rho_{\mathrm{out}}(\rho)&=\frac{K\rho K^\dagger}{S}\\\nonumber
  &=\frac{1}{S}\sum_j p_j\,K|\phi_j\rangle\!\langle\phi_j|K^\dagger\\\nonumber
  &=\frac{1}{S}\sum_j p_j\,|\psi_j\rangle\!\langle\psi_j|\\\nonumber
  &=\frac{1}{S}\sum_{j:\,s_j>0} p_j\,s_j\,|\varphi_j\rangle\!\langle\varphi_j|\\
  &=\sum_{j:\,s_j>0}\frac{p_j s_j}{S}\,|\varphi_j\rangle\!\langle\varphi_j|.
  \end{align}

Concurrence is convex on two-qubit states, hence
  \begin{align}
C\bigl(\rho_{\mathrm{out}}(\rho)\bigr)\;\le\;\sum_{j:\,s_j>0}\frac{p_j s_j}{S}\,C_j.
  \end{align}
  Multiplying by $S$ and adding $S$,
\begin{widetext}
  \begin{align}
  S\Bigl(1+C\bigl(\rho_{\mathrm{out}}(\rho)\bigr)\Bigr)\;\le\;S+\sum_j p_j s_j C_j
  \;=\;\sum_j p_j\bigl(s_j+s_j C_j\bigr)\;\le\;\sum_j p_j\cdot 1\;=\;1,
  \end{align}
\end{widetext}
where the last inequality uses $s_j(1+C_j)\le 1$ termwise.
If $S=0$, the inequality is trivial. Since $S=P_{\mathrm{succ}}(\rho)$, we arrive at:
\begin{align}
P_{\mathrm{succ}}(\rho)\,\bigl[1+C\bigl(\rho_{\mathrm{out}}(\rho)\bigr)\bigr]\;\le\;1.
\end{align}
\begin{figure*}[t]
  \includegraphics[width=1.0\textwidth]{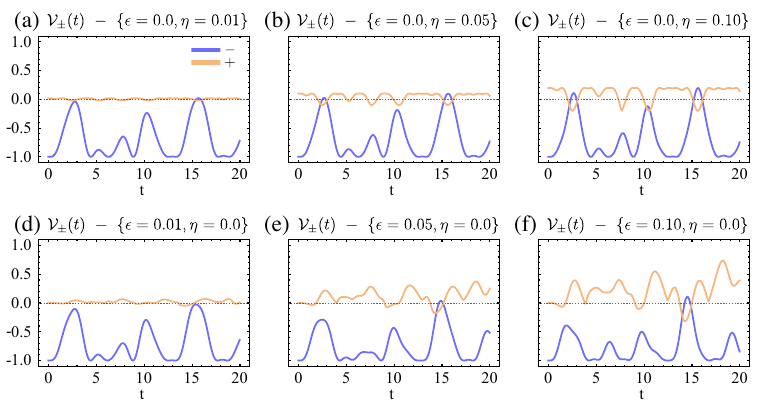}
  \centering
  \caption{Robustness of the trade-off under different conditions of weak initial entanglement and weak interactions. Each panel shows the time dependence of $\mathcal{V}_\pm(t) = P_{\mathrm{succ},\pm}(t)\bigl(1 + C_\pm(t)\bigr) - 1$ for the $|-\rangle^C$ (blue) and $|+\rangle^C$ (yellow) branches. Top row: non-interacting dynamics ($\epsilon = 0$) with increasing initial entanglement $\eta$ in the state $\lvert\psi(\eta)\rangle = \bigl(\lvert 00\rangle + \eta \lvert 11\rangle\bigr)/\sqrt{1+\eta^2}$, for (a) $\eta = 0.01$, (b) $\eta = 0.05$, and (c) $\eta = 0.10$. Bottom row: separable input ($\eta = 0$) with increasing interaction strength in the isotropic coupling $\mathcal{H}_{\mathrm{int}}$, for (d) $\epsilon = 0.01$, (e) $\epsilon = 0.05$, and (f) $\epsilon = 0.10$. In all cases with small $\eta$ ($\epsilon$), $\mathcal{V}_\pm(t)$ remains close to zero, and the amplitude of positive excursions grows smoothly with $\eta$ or $\epsilon$, illustrating that the trade-off is structurally stable, that is small deviations from the ideal assumptions lead only to small, continuous deviations from the ideal bound $P_{\mathrm{succ}}(1 + C)\le 1$.}
  \label{fig:weak_2}
\end{figure*}

\section{A Numerical Study of the Robustness of the Bound}
\label{sec:perturb}
In this Appendix we provide a numerical study of how the trade-off relation behaves when the two structural assumptions, that is conditions of non–interacting dynamics and separable input, are \emph{weakly} violated. We focus on the same Hamiltonians $\mathcal{H}_M$ and $\mathcal{H}_N$ (specifically, we assume $\omega_A=1.0, \omega_B=1.0, c_{M,A}^X=1.0, c_{N,B}^X=1.0$) as in Sec.~\ref{sec:5} and consider the following perturbations:

\begin{enumerate}
  \item \textit{Weak initial entanglement.}
  The target qubits are initialized in
\begin{align}
    \lvert \psi(\eta)\rangle = \frac{1}{\sqrt{1+\eta^2}}\bigl(\lvert 00\rangle + \eta \lvert 11\rangle\bigr),
\end{align}
where $\eta\in\mathbb{R},\ |\eta|\ll 1$, with no interaction between them ($\epsilon = 0$).

  \item \textit{Weak cross–talk.}
  The target qubits start from the separable state $\lvert 00\rangle$, but each process Hamiltonian is modified by the addition of an isotropic interaction term
\begin{align}
     \mathcal{H}_k \;&\to\; \mathcal{H}_k + \epsilon \mathcal{H}_{\mathrm{int}},\\\qquad
     \mathcal{H}_{\mathrm{int}} &= \sigma_x\otimes\sigma_x + \sigma_y\otimes\sigma_y + \sigma_z\otimes\sigma_z,
\end{align}
  with small coupling $\epsilon > 0$ and $k\in\{M,N\}$.
\end{enumerate}

For each branch of the control measurement ($\pm$) we compute the success probability $P_{\mathrm{succ},\pm}(t)$, the concurrence $C_\pm(t)$, and the ``violation'':
\begin{align}
  \mathcal{V}_\pm(t) := P_{\mathrm{succ},\pm}(t)\bigl(1 + C_\pm(t)\bigr) - 1.
\end{align}
The trade-off proven in the main text implies $\mathcal{V}_\pm(t)\le 0$ for strictly non-interacting dynamics and separable inputs.
Here we use $\mathcal{V}_\pm(t)$ as a diagnostic to quantify deviations from the ideal assumptions.

Fig.~\ref{fig:weak_1} shows $\mathcal{V}_\pm(t)$ together with $C_\pm(t)$ and $P_{\mathrm{succ},\pm}(t)$ for three cases:
(a) the baseline protocol with $\epsilon=0$, $\eta=0$;
(b) weak initial entanglement with $\epsilon=0$, $\eta=0.05$; and
(c) weak cross–talk with $\epsilon=0.05$, $\eta=0$.
The inequality is always satisfied in (a), with $\mathcal{V}_\pm(t)\le 0$ at all times.
In contrast, in panel (b) and (c) small positive excursions of the violation function become visible when $\eta>0$ or $\epsilon>0$.

A more systematic picture is provided in Fig.~\ref{fig:weak_2}, where we plot $\mathcal{V}_\pm(t)$ for different values of $\eta$ at $\epsilon=0$ [panels~(a-c)] and for different values of $\epsilon$ at $\eta=0$ [panels~(d-f)].
These numerical results indicate that the trade–off relation is robust in a continuity sense: small deviations from the ideal assumptions (either through weak initial entanglement or weak interactions) lead only to small deviations of $\mathcal{V}_\pm(t)$ from zero.
At the same time, they confirm that once genuine interactions are present, even if weak, the strict universal bound $P_{\mathrm{succ}}(1+C)\le 1$ no longer holds in general, as expected on physical grounds.

\section{Derivations of the Conditions for Constant Maximal-Concurrence Generation}
\label{sec:der_2021}

In the following, we provide an explicit derivation for the results presented in Eq.~\eqref{eq:rhoabentries_ijkl} and Eq.~\eqref{eq:defg}. We demonstrate how the specific Hamiltonian symmetries lead to a time-independent, maximally entangled state for the $|-\rangle^C$ post-selection branch.

We begin with the initial separable state $|\psi_\text{sep}\rangle = |00\rangle^{AB}$, which corresponds to Cases $\mathbb{A}$ and $\mathbb{B}$ in Table~\ref{tab:table1}. The unnormalized post-selected state is given by the action of the commutator on this initial state:
\begin{align}
	\ket{\widetilde{\Psi}}^{-,AB}  = \frac{1}{2}(U_{\mathcal{M}}U_{\mathcal{N}} - U_{\mathcal{N}}U_{\mathcal{M}})|00\rangle^{AB}.	
\end{align}

This state vector is equivalent to the first column of the $4 \times 4$ operator matrix 
\begin{align}
\label{eq:deltU_mn}
	\Delta U(t) = U_{\mathcal{M}}(t/2)U_{\mathcal{N}}(t/2) - U_{\mathcal{N}}(t/2)U_{\mathcal{M}}(t/2). 	
\end{align}

The next task is therefore to explicitly calculate the four components of this column vector. The Hamiltonians for these cases are defined in Sec.~\ref{sec:5} as $\mathcal{H}_M = \mathcal{H}_{\text{int}} + \mathcal{H}_{\text{aux},M}^{(A)}$ and $\mathcal{H}_N = \mathcal{H}_{\text{int}} + \mathcal{H}_{\text{aux},N}^{(B)}$, where $\mathcal{H}_{\text{int}}$ is the internal Hamiltonian (with $\omega_A = \omega_B = \omega_z$) and the auxiliary terms are $\mathcal{H}_{\text{aux},M}^{(A)} = c_{M,A}^X \sigma_x^A$ and $\mathcal{H}_{\text{aux},N}^{(B)} = c_{N,B}^X \sigma_x^B$.

The unitaries are given by $U_{\mathcal{M}}(t/2) = \exp[-i\mathcal{H}_M (t/2)]$ and $U_{\mathcal{N}}(t/2) = \exp[-i\mathcal{H}_N (t/2)]$. By explicitly computing according to Eq.~\eqref{eq:deltU_mn} we find the four components of the first column of $\Delta U(t)$ are 
\begin{widetext}
\begin{align}
	&\bra{00}^{AB}\Delta U(t)\ket{00}^{AB}=0\\\nonumber
	&\bra{01}^{AB}\Delta U(t)\ket{00}^{AB}=\\
&\frac{
2 c_{N,B}^X e^{-\frac{i}{2}\omega_z t}
\sin\!\left(\frac{\omega_z t}{2}\right)
\left[
\sqrt{(c_{M,A}^X)^2 + \omega_z^2}\,
\cos\!\left(\frac{t}{2}\sqrt{(c_{M,A}^X)^2 + \omega_z^2}\right)
- i \omega_z
\sin\!\left(\frac{t}{2}\sqrt{(c_{M,A}^X)^2 + \omega_z^2}\right)
\right]
\sin\!\left(\frac{t}{2}\sqrt{(c_{N,B}^X)^2 + \omega_z^2}\right)
}{
\sqrt{(c_{M,A}^X)^2 + \omega_z^2}\,
\sqrt{(c_{N,B}^X)^2 + \omega_z^2}
}\\\nonumber
	&\bra{10}^{AB}\Delta U(t)\ket{00}^{AB}=\\
&\frac{
2 c_{M,A}^X e^{-\frac{i}{2}\omega_z t}
\sin\!\left(\frac{\omega_z t}{2}\right)
\left[
-\sqrt{(c_{N,B}^X)^2 + \omega_z^2}\,
\cos\!\left(\frac{t}{2}\sqrt{(c_{N,B}^X)^2 + \omega_z^2}\right)
+ i \omega_z
\sin\!\left(\frac{t}{2}\sqrt{(c_{N,B}^X)^2 + \omega_z^2}\right)
\right]
\sin\!\left(\frac{t}{2}\sqrt{(c_{M,A}^X)^2 + \omega_z^2}\right)
}{
\sqrt{(c_{M,A}^X)^2 + \omega_z^2}\,
\sqrt{(c_{N,B}^X)^2 + \omega_z^2}
}\\
	&\bra{11}^{AB}\Delta U(t)\ket{00}^{AB}=0
\end{align}
\end{widetext}
From the above, one can easily observe that when $ c_{M,A}^{X} = \pm c_{N,B}^{X}$, or the ratio $\mathcal{R} = c_{M,A}^{X} / c_{N,B}^{X}$ is set to $\pm 1$, the specific symmetries of the Hamiltonian cause an alignment of the amplitude of $\bra{01}^{AB}\Delta U(t)\ket{00}^{AB}$ and $\bra{10}^{AB}\Delta U(t)\ket{00}^{AB}=$.
As a consequence, under the condition, e.g., in the case $\mathcal{R} = -1$ where the coefficients of $\ket{01}^{AB}$ and $\ket{10}^{AB}$ are identical, up to normalization, the resulting state is recognized as the fully entangled $\frac{1}{\sqrt{2}}(|01\rangle^{AB} + |10\rangle^{AB})$ state. The concurrence is therefore $\mathcal{C}_- = 1$ for all $t$. This same logic applies to the case $\mathcal{R} = 1$, yielding the state $\frac{1}{\sqrt{2}}(|01\rangle^{AB} - |10\rangle^{AB})$. Furthermore, because we have the following relationship:
\begin{align}
	\widetilde{\rho}_{AB}^{-}=\ketbra{\widetilde{\Psi}}^{-,AB},
\end{align}
the components that have finite value in $\widetilde{\rho}_{AB}^{-}$ are made from non-zero entries in $\ket{\widetilde{\Psi}}^{-,AB}$, which are associated with the following terms,
\begin{align}
	\ketbra{01}{01}^{-,AB},\\
	\ketbra{10}{10}^{-,AB},\\
	\ketbra{01}{10}^{-,AB},\\
	\ketbra{10}{01}^{-,AB}.
\end{align}
By a direct calculation with basic algebraic manipulations, we arrive at, for $\mathcal{R}=-1$, 
\begin{widetext}
\begin{align}
	\bra{01}^{AB}\widetilde{\rho}_{AB}^{-}\ket{01}^{AB}&={c_{M,A}^X}^2\sin^2({\omega_z  t/2})\sin({\Theta  t})\frac{(2\omega_z^2+{c_{M,A}^X}^2/2)\sin({\Theta  t})+({c_{M,A}^X}^2/2)\sin{(3  \Theta  t)}}{32\Theta^4}\\
	\bra{10}^{AB}\widetilde{\rho}_{AB}^{-}\ket{10}^{AB}&={c_{M,A}^X}^2\sin^2({\omega_z  t/2})\sin({\Theta  t})\frac{(2\omega_z^2+{c_{M,A}^X}^2/2)\sin({\Theta  t})+({c_{M,A}^X}^2/2)\sin{(3  \Theta  t)}}{32\Theta^4}\\
	\bra{01}^{AB}\widetilde{\rho}_{AB}^{-}\ket{10}^{AB}&={c_{M,A}^X}^2\sin^2({\omega_z  t/2})\sin({\Theta  t})\frac{(2\omega_z^2+{c_{M,A}^X}^2/2)\sin({\Theta  t})+({c_{M,A}^X}^2/2)\sin{(3  \Theta  t)}}{32\Theta^4}\\
	\bra{10}^{AB}\widetilde{\rho}_{AB}^{-}\ket{01}^{AB}&={c_{M,A}^X}^2\sin^2({\omega_z  t/2})\sin({\Theta  t})\frac{(2\omega_z^2+{c_{M,A}^X}^2/2)\sin({\Theta  t})+({c_{M,A}^X}^2/2)\sin{(3  \Theta  t)}}{32\Theta^4}
\end{align}
\end{widetext}
while for $\mathcal{R}=1$,
\begin{widetext}
\begin{align}
	\bra{01}^{AB}\widetilde{\rho}_{AB}^{-}\ket{01}^{AB}={c_{M,A}^X}^2\sin^2({\omega_z  t/2})\sin({\Theta  t})\frac{(2\omega_z^2+{c_{M,A}^X}^2/2)\sin({\Theta  t})+({c_{M,A}^X}^2/2)\sin{(3  \Theta  t)}}{32\Theta^4}\\
	\bra{10}^{AB}\widetilde{\rho}_{AB}^{-}\ket{10}^{AB}=-{c_{M,A}^X}^2\sin^2({\omega_z  t/2})\sin({\Theta  t})\frac{(2\omega_z^2+{c_{M,A}^X}^2/2)\sin({\Theta  t})+({c_{M,A}^X}^2/2)\sin{(3  \Theta  t)}}{32\Theta^4}\\
	\bra{01}^{AB}\widetilde{\rho}_{AB}^{-}\ket{10}^{AB}=-{c_{M,A}^X}^2\sin^2({\omega_z  t/2})\sin({\Theta  t})\frac{(2\omega_z^2+{c_{M,A}^X}^2/2)\sin({\Theta  t})+({c_{M,A}^X}^2/2)\sin{(3  \Theta  t)}}{32\Theta^4}\\
	\bra{10}^{AB}\widetilde{\rho}_{AB}^{-}\ket{01}^{AB}={c_{M,A}^X}^2\sin^2({\omega_z  t/2})\sin({\Theta  t})\frac{(2\omega_z^2+{c_{M,A}^X}^2/2)\sin({\Theta  t})+({c_{M,A}^X}^2/2)\sin{(3  \Theta  t)}}{32\Theta^4}
\end{align}
\end{widetext}
As a result, if we define
\begin{align}\nonumber
	\mathcal{G}=&{c_{M,A}^X}^2\sin^2({\omega_z  t/2})\sin({\Theta  t})\times\\
	&\frac{(2\omega_z^2+{c_{M,A}^X}^2/2)\sin({\Theta  t})+({c_{M,A}^X}^2/2)\sin{(3  \Theta  t)}}{16\Theta^4}
\end{align}
one can write the (unnormalized) density operator for the $\ket{-}$ branch post-selection in a compact form as shown below,
\begin{align}
	\label{eq:rhoabentries}
	\bra{ij}^{AB}\widetilde{\rho}_{AB}^{-}\ket{kl}^{AB}=
    \begin{cases}
      \mathcal{G}/2 & \text{$(i,k=s$ and $j,l=1-s)$}\\
      -\mathcal{R}\mathcal{G}/2 & \text{$(i,l=s$ and $j,k=1-s)$}\\
      0 & \text{otherwise}.
    \end{cases}
\end{align}
where $s=0$ or $1$, which is the same form as we presented in Eq.~\eqref{eq:defg}, and $\mathcal{G}$ is recognized to be the corresponding success probability function. These analysis holds for other choices of $|\psi_\text{sep}\rangle$, therefore leading to our conclusion presented in Table~\ref{tab:table1}.

\end{document}